\documentclass[manuscript,screen]{acmart}
\usepackage{graphicx}
\usepackage{subfig}
\usepackage{xspace}
\usepackage{booktabs}
\usepackage{enumitem}

\usepackage{amsmath,amssymb}
\usepackage[bb=boondox,bbscaled=.95,cal=boondoxo]{mathalfa}
\newcommand{\X}{$\mathbb{X}$\xspace}
\AtBeginDocument{%
  }

\setcopyright{acmlicensed}
\copyrightyear{2018}
\acmYear{2018}
\acmDOI{XXXXXXX.XXXXXXX}
\acmConference[Conference acronym 'XX]{Make sure to enter the correct
  conference title from your rights confirmation email}{June 03--05,
  2018}{Woodstock, NY}
\acmISBN{978-1-4503-XXXX-X/2018/06}

\begin{document}

\title{Needs Your Help: Understanding Platform-Directed Rating Participation in Community Notes on X}

\author{Shuning Zhang}
\email{zsn23@mails.tsinghua.edu.cn}
\author{Changxi Wen}
\email{wcx24@mails.tsinghua.edu.cn}
\affiliation{
    \institution{Tsinghua University}
    \city{Beijing}
    \country{China}
}
\author{Jiuchang Wang}
\affiliation{
    \institution{Communication University of China}
    \city{Beijing}
    \country{China}
}
\author{Dai Shi}
\affiliation{
    \institution{Tongji University}
    \city{Shanghai}
    \country{China}
}
\author{Gabriele Lenzini}
\author{Yuwei Chuai}
\authornotemark[1]
\affiliation{
    \institution{University of Luxembourg}
    \city{Esch-sur-Alzette}
    \country{Luxembourg}
}
\author{Xin Yi}
\authornote{Corresponding author.}
\affiliation{
    \institution{Tsinghua University}
    \city{Beijing}
    \country{China}
}

\renewcommand{\shortauthors}{Trovato et al.}

\begin{abstract}
Community-based fact-checking is promising in countering misinformation, yet its scalability is constrained by slow rating accumulation. To address this challenge, platforms such as X implement platform-directed rating mechanisms, specifically through ``Needs Your Help'' algorithmic prompts, to target unresolved notes. Using a dataset of over 220 million rating contributions -- including 1.9 million platform-directed ratings -- on X, we examine contributors' response to note prompts, notes' resolution, and raters' spillovers. We found (i) at note level, population-sampled ratings concentrate on recent notes with certain helpfulness and high disagreement. Once sampled, population-sampled rating was associated with faster and more transitions to resolved statuses. (ii) At rater level, following raters' first observed population-sampled rating, raters exhibit significant yet modest increases in daily ratings, rating pace and tag usage, while other behaviors show no change. These highlight the promise of algorithmic nudges to guide volunteer attention toward contested content, accelerating consensus while sustaining rater engagement.
\end{abstract}

\ccsdesc[300]{Human-centered computing~Empirical studies in collaborative and social computing}
\ccsdesc[500]{Human-centered computing~Social media}
\ccsdesc[300]{Security and privacy~Social aspects of security and privacy}

\keywords{Misinformation, Social media, Community Notes, Crowdsourced fact-checking}

\begin{teaserfigure}
 \includegraphics[width=\textwidth]{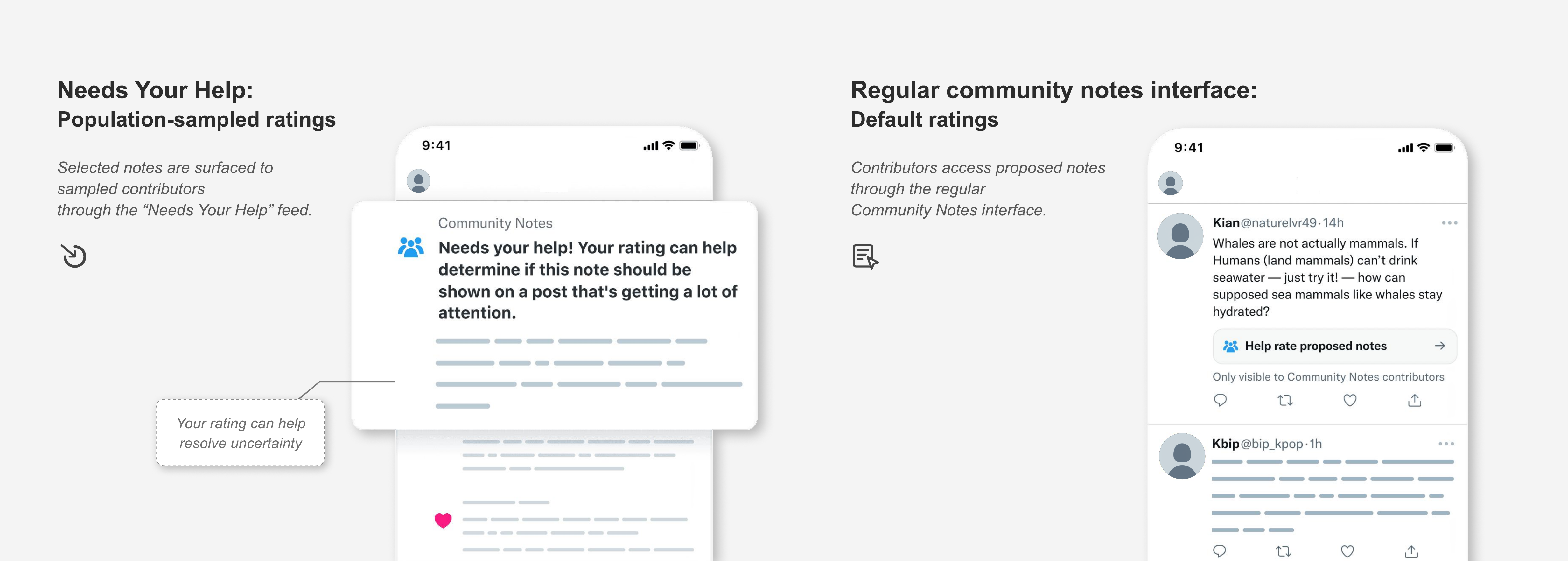}
  \caption{Illustration of platform-directed ``Needs Your Help'' notification (left panel) and regular rating access channel (right panel) for contributors in the Community Notes program on \X.}
  \Description{Illustration of two rating pathways in the Community Notes system. On the left, selected community notes are surfaced to sampled contributors through the 'Needs your help' feed. On the right, a post in the regular timeline includes a 'Help rate proposed notes' entry through which contributors can access notes for default ratings.}
  \label{fig:teaser}
\end{teaserfigure}


\maketitle

\section{Introduction}

The viral spread of misleading information on social media, alongside regulatory obligations to mitigate systemic risks under frameworks such as the EU's Digital Services Act~\cite{wilman2024eu}, underscores the need for scalable fact-checking approaches~\cite{micich2023misinformation,xu2025misinformation,chuai2026community,ecker2024misinformation}. One such promising approach is community-based fact-checking, exemplified by the Community Notes program that originated on \X and expanded to other major social media platforms, including Meta~\cite{meta_community_notes}, TikTok~\cite{presser2025footnotes}, and YouTube Shorts~\cite{youtube2024context}.
Community-based fact-checking enables volunteer contributors to write contextual notes on potentially misleading posts and evaluate the helpfulness of community notes written by others~\cite{prollochs2022community,wojcik2022birdwatch}. A bridging algorithm uses these evaluations to identify helpful notes that are supported by cross-perspective contributors and feature them for public display~\cite{communitynotes_ranking_algorithm}. Research shows that displayed community notes can effectively reduce the engagement with and the spread of misleading posts~\cite{chuai2026community,chuai2024did}. However, delays in note publication can limit opportunities to intervene at the early viral stages where posts accumulate substantial engagement~\cite{chuai2024did}. Therefore, gathering timely evaluations is crucial for translating contributors' proposed corrections into publicly available context.

Gathering these evaluations presents participation and coordination challenges. In an environment shaped by personalized social media feeds, contributors choose which notes to evaluate from those they encounter, making assessment progress dependent on both the availability of relevant notes and contributors' willingness to rate them. Prior Human-Computer Interaction (HCI) research has shown that unguided crowds struggle to focus on high-priority tasks without system scaffolding~\cite{deng2023understanding}. As a remedy, targeted task routing based on users' interests is effective at boosting engagement in communities such as Wikipedia~\cite{cosley2007suggestbot}, while requests that highlight the distinct value of a user's contribution can further encourage participation~\cite{beenen2004using}. Together, these findings motivate examining how platforms actively connect contributors with tasks that need their attention within community-based fact-checking systems. 

On \X, the Community Notes program includes a ``Needs Your Help'' feature which pushes notifications to randomly sampled contributors, inviting them to rate selected notes with the aim of fostering consensus building and accelerating note publication (Figure~\ref{fig:teaser}). Specifically, the system identifies candidate notes written on recent, high-visibility posts that are close to reaching a consensus threshold, particularly those labeled as ``Needs More Ratings'' with positive signals or lopsided rating distributions. Eligible notes are surfaced through a dedicated feed tab, contextual timeline previews and push notifications sent to a randomized subset of eligible raters, soliciting evaluations from diverse perspectives. This mechanism combines platform selection of notes, sampling of contributors, and contributors' voluntary decisions to respond. Consequently, observed participation reflects both the platform's targeting and contributors' responses. Although the selection criteria are documented, they provide limited insight into which notes ultimately receive completed ratings and how this participation relates to subsequent outcomes. Thus, the observed rating records provide a unique opportunity to study where this platform-directed rating participation occurs in practice and how it relates to the progression of collective evaluations and contributors' subsequent activity.

\vspace{.5em}
\noindent \textbf{Research questions.}
In this study, we examine the ``Needs Your Help'' feature on \X, an intervention designed to prompt contributors to evaluate specific notes. While presented with platform prompts, contributors retain the autonomy to choose which notes to evaluate. Therefore, their participation may shift note evaluation dynamics and accelerate consensus. Beyond note outcomes, receiving such prompts may also reshape contributors' own ongoing engagement. Accordingly, we investigate both note-level and rater-level effects through two research questions (RQs):
\begin{itemize}[leftmargin=*,noitemsep,topsep=0pt]
    \item \textbf{RQ1 (Note-level).} Which note characteristics and rating dynamics are associated with the receipt of population-sampled ratings, and how is this intervention associated with note resolution speed and status?
    \item \textbf{RQ2 (Rater-level).} How does raters' evaluation activity change after their first population-sampled rating?
\end{itemize}

\vspace{.5em}
\noindent \textbf{Data and methods.} To address these research questions, we analyze $\mathbb{X}$'s public Community Notes dataset, containing 220 million ratings on 2.8 million notes, and 1.9 million population-sampled ratings (i.e., platform-directed participation). For RQ1, we compare the pre-event characteristics of notes that have received population-sampled ratings with matched notes that never receive such ratings by estimating a multivariate binary logistic regression model on post-level, note-level and pre-event rating characteristics. We then construct matched control group of notes not covered by population-sampled rating using Propensity Score Matching (PSM), and use Difference-in-Differences (DID) analysis to how receiving population-sampled ratings affects note status resolution speed, and transition probabilities from \textit{``Needs More Ratings''} (NMR) to \textit{``Currently Rated Helpful''} (CRH) or \textit{``Currently Rated Not Helpful''} (CRNH). For RQ2, we employ a staggered DID design to compare contributors' behaviour before they submit their first population-sampled rating with their behaviour afterward, and examine how these changes vary across contributor characteristics.

\vspace{.5em}
\noindent \textbf{Contributions.} Collectively, this paper contributes: 
(i) the first comprehensive analysis of population-sampled rating mechanisms, helping Human-Computer Interaction (HCI) researchers to understand how platform algorithms steer crowd attention toward disputed content; 
(ii) empirical evidence that targeted prompts accelerate consensus without causing volunteer fatigue, showing for the HCI community how nudges can scaffold high-quality feedback while sustaining voluntary participation; 
(iii) design implications that guide HCI practitioners to build dedicated task feeds that reduce bias and optimize crowd task allocation.



\section{Background and Related Work}

Decentralized fact-checking relies on crowd consensus, algorithmic routing, and curated presentation to deliberate on disputed information. In this section, we review four related areas: (i) crowd-sourced rating, (ii) algorithmic nudging, (iii) targeted feeds curation, and (iv) the ecosystem of Community Notes. 

\subsection{Crowdsourced Rating}
Crowd-based evaluation provides a scalable alternative to expert-based fact-checking by distributing verification tasks across online users. Prior studies show that while individual user judgments are often noisy, aggregating assessments from politically balanced groups can approximate expert fact-checker benchmarks~\cite{allen2021scaling}. However, crowd performance varies substantially across task designs, participant demographics, political leanings, and article types~\cite{bhuiyan2020investigating}. On community-driven forums like Reddit's r/DebunkThis, contributors debunk false claims through diverse strategies, such as challenging source credibility, citing authoritative external evidence, offering step-by-step corrections, and engaging in collaborative reasoning~\cite{he2022help}. Similar collaborative annotation designs, such as spatial-temporal bounding algorithms for deepfake videos, likewise aggregate distributed user inputs to verify multimodal content~\cite{zhang2026collab}. 

Despite these capabilities, voluntary crowd rating faces practical challenges. Laypeople generally rate mainstream sources as more trustworthy than hyperpartisan sources, though trust diverges across ideologies~\cite{pennycook2019fighting}. Consequently, early Twitter Birdwatch trials revealed noticeable partisan asymmetries in target selection and content evaluation~\cite{allen2022birds}, showing that raw crowd ratings may not yield actionable consensus~\cite{saeed2022crowdsourced}. In response, platforms deployed bridging-based consensus algorithms that weigh historical rating diversity to elevate notes with cross-perspective agreement~\cite{wojcik2022birdwatch}. Note characteristics strongly influence this consensus. Citations from credible external sources and an emotionally restrained writing style increase perceived helpfulness, whereas hyperpartisan sources or notes targeting political figures face resistance~\cite{prollochs2022community,solovev2025references,yao2024readable}. 

Nonetheless, organic voluntary participation faces challenges. Volunteer contributors and professional fact-checkers prioritize different claims~\cite{pilarski2024community}, and professional fact-checkers themselves show variations in political topic selection~\cite{chuai2025fact}. Crucially, organically submitted notes frequently experience severe latency, failing to reach consensus before misinformation achieves viral diffusion~\cite{chuai2024did,chuai2026community}. While computational simulations like MultiCom~\cite{wen2026towards} and predictive models like COMMUNITYNOTES~\cite{xing2026spatiotemporal} forecast potential rating trajectories, they do not examine how real-world platform routing directs human attention to content. \textbf{\textit{Unlike prior studies that analyze unguided organic contributions, our work empirically investigates how platform-driven sampling coordinates human raters to resolve contentious notes.}}

\subsection{Algorithmic Nudging}

Algorithmic nudging shapes online behaviour by altering the visibility, ordering, and accessibility of available actions without banning individual options~\cite{weinmann2016digital}. Modern recommender systems function as digital choice architectures, determining the specific items and tasks users encounter~\cite{jesse2021digital}. In crowdsourcing environments, micro-interventions such as push notifications and task recommendations steer contribution rates and system outputs~\cite{kim2016analysis}. However, poorly calibrated nudges can introduce negative externalities. For instance, excessive or poorly timed prompts risk triggering volunteer fatigue and contributor attrition~\cite{bashirieh2017nudge}. These dynamics highlight the tension between platform steering and voluntary labor sustainability.

In crowdsourced governance, intervention frameworks like the six-dimension model proposed by Lloyd et al.~\cite{lloyd2026beyond} emphasize that design choices across participation, curation, and presentation carry profound normative implications for platform deliberation. Recent fact-checking interventions primarily targeted note generation and user credibility assessment. For example, note request features allow users to flag misleading posts, yet only a small fraction of flagged content receives actionable contributor attention~\cite{chuai2026request}. Tailored credibility feedback helps users calibrate their assessments of ambiguous claims~\cite{tang2025between}, while collaborative annotation scaffolds guide attention toward pivotal evidence for manipulated media~\cite{zhang2026collab}. Other systems actively support draft generation. Commenotes integrates organic, real-time responses into early note drafts~\cite{zhang2025commenotes}, CANote facilitates source retrieval and draft authoring via human-AI collaboration~\cite{zhang2026canote}, and Supernotes synthesizes multi-author perspectives to foster broader consensus~\cite{de2025supernotes}. \textbf{\textit{While existing interventions primarily optimize how notes are authored or presented, our work focuses on the downstream evaluation, specifically evaluating how algorithmic prompts nudge raters toward contested notes and whether this intervention induces spillover or volunteer fatigue.}}

\subsection{Targeted Feeds Curation}

The architecture through which content and tasks are surfaced governs user engagement and cognitive attention. Users actively construct internal mental representations, termed ``folk theories'', to rationalize how curation algorithms filter and prioritize content~\cite{eslami2016first} instead of passively consuming static feeds. When interfaces make algorithmic seams visible, as shown in decentralized platforms like Mastodon, users build higher trust and actively negotiate trade-offs between machine learning and heuristic filtering rules~\cite{liu2025understanding}. Nevertheless, misalignments frequently surface between user mental models and platform curation mechanisms. For example, many users remain unaware of feed personalization and ad-targeting controls, misunderstanding how their feeds are constructed~\cite{hsu2020awareness}. 

To bridge this gap, HCI research has increasingly explored feed curation experiences~\cite{feng2024mapping}. Popowski et al.~\cite{popowski2026social} introduced feed elicitation interviews to help users translate nuanced personal preferences into customized feeds, showing that structured preference elicitation outperforms manual feed curation. Similarly, BONSAI enables users to specify natural language intents and exercise transparent control over their feeds~\cite{el2026bonsai}. Piccardi et al.~\cite{piccardi2026reranking} showed that real-time feed reranking via browser extensions proves powerful to evaluate real-world informational interventions. Epstein et al.~\cite{epstein2026value} revealed that algorithmic feed amplification often shows negative correlations with users' explicit personal values, selectively boosting moral expressions. \textbf{\textit{While prior feed curation literature focuses on user-facing personalization, algorithmic transparency, and consumption feeds, we investigated targeted task feeds deployed by platforms to direct debunking.}}

\subsection{Community Notes}

Community Notes (formerly Birdwatch) on \X represents one of the largest deployments of crowdsourced debunking~\cite{x_community_notes}. Qualified contributors write contextual notes on potentially misleading posts, while qualified raters evaluate whether those notes are ``Helpful,'' ``Somewhat Helpful,'' or ``Not Helpful''~\cite{wojcik2022birdwatch}. Instead of relying on raw majority voting, the platform implements a matrix-factorization bridging algorithm that scores notes based on whether they receive positive ratings across raters who historically hold divergent perspectives~\cite{x_note_ranking_algorithm}. Once a note achieves a sufficiently high consensus score, it receives a ``Currently Rated Helpful'' status and becomes publicly displayed on the associated post.

Studies confirm that publicly displayed community notes reduce engagement with and retweets of misleading posts~\cite{chuai2026community}. However, the platform's participation model introduces bottlenecks. Many notes reach CRH status only after the underlying post has garnered substantial engagement, severely limiting the intervention's efficacy~\cite{chuai2024did}. Moreover, the introduction of public notes can occasionally polarize community discussions, provoking heightened moral outrage and negative sentiment in post replies~\cite{chuai2025community}. 

To accelerate deliberation on unresolved notes, \X introduced ``Needs Your Help'' notifications, which sample specific contributors and prompt them to evaluate pending notes~\cite{x_timeline_tabs}. Ratings collected through this channel are logged as ``population-sampled ratings.'' However, empirical questions remain unanswered around what specific note attributes trigger population sampling, whether sampled ratings accelerate status resolution from NMR to a definitive state, and whether completing these prompted tasks alters contributors' subsequent evaluation. \textbf{\textit{While earlier research evaluates algorithmic performance, our work provides an end-to-end empirical analysis of the population-sampled rating mechanism at the note, rating, and rater levels.}}

\section{Methodology}

\subsection{Dataset}

We downloaded the public archive of the Community Notes dataset on \X, covering January 23, 2021 to June 15, 2026 (UTC)\footnote{https://communitynotes.x.com/guide/en/under-the-hood/download-data}. It contained 220,472,232 ratings for 2,859,500 notes with at least one rating. Among these ratings, 218,537,338 were \textit{DEFAULT} ratings and 1,934,894 came from population sampling, as indicated by the rating source labels provided by \X. Because the population sampling program began on February 1, 2025, we restricted our analysis to notes created on or after this date. The resulting sample contained 93,878,870 ratings, including 91,948,929 \textit{DEFAULT} ratings and 1,929,941 ratings from population sampling. For subsequent analyses, we calculated helpfulness scores using \X's official mapping: \textsc{Helpful} was coded as 1, \textsc{Somewhat Helpful} as 0.5, and \textsc{Not Helpful} as 0.
Among the notes in our analysis sample, 159,377 received at least one population-sampled rating, accounting for 12.95\% of all notes, while 1,071,048 notes had not received any population-sampled rating prior to our data cutoff. At the time of the first population-sampled rating, 95.6\% of the corresponding notes remained in the NMR status, and they had received a median of 15 prior DEFAULT ratings. A median of 1.3 hours elapsed between note creation and the first population-sampled rating.

\subsection{Prevalence (RQ1)}

We first examined which notes are more likely to receive population-sampled ratings. 

\subsubsection{Estimation Model and Factors}

We estimated a binary logistic regression model where the outcome represents observed population-sampled rating (denoted POP) entry. The predictors capture post-level, note-level, and pre-event rating characteristics. All three groups of predictors were included in the regression model.

\textit{Post-level characteristics.}
Prior research on Birdwatch examined the characteristics of the posts being fact-checked, including whether they were classified as misleading and the reasons selected by note authors~\cite{prollochs2022community}. Referring to the prior work, we examined post classifications and author-selected reasons. Additionally, we considered \textit{the delay between post publication and note creation}, and \textit{whether the same post already had a note}, to capture the timing and existing annotation context associated with observed population-sampled rating.

\textit{Note-level characteristics.}
Prior studies have demonstrated that note content and external sources were strongly associated with perceived helpfulness~\cite{prollochs2022community,solovev2025references}. We therefore examined \textit{note text length}, \textit{URL count}, and \textit{the author-selected trustworthy-sources flags}. We additionally considered \textit{note format and age} to capture differences in the type and lifecycle stages of notes. We also examined the note \textit{author's prior record}, as \X's contributor scoring framework weighs the helpfulness of previously authored notes\footnote{\url{https://communitynotes.x.com/guide/en/under-the-hood/contributor-scores}}.

\textit{Pre-event rating characteristics.}
Prior research has examined changes in rating volume and rating leaning around note display~\cite{chuai2026consensus}, while research on note requests has considered both helpfulness and polarization when comparing notes~\cite{chuai2026request}. Therefore, we examined three aspects of the pre-event rating history: \textit{rating activity}, \textit{existing Helpful support}, and \textit{dispersion across rating categories}. We considered DEFAULT rating counts across pre-event time windows to distinguish recent activity from earlier activity, and used prior Helpful mean and rating entropy to summarize existing support and rating dispersion, respectively. We then examined whether these characteristics were associated with observed population-sampled rating.
Prior work has used normalized entropy to quantify disagreement among annotations~\cite{ramas2021identifying}. We calculated rating entropy as
\[
D_i=-\frac{\sum_{k=1}^{3}p_{ik}\log p_{ik}}{\log 3},
\]
where \(p_{ik}\) is the proportion of pre-event \textit{DEFAULT} ratings in category \(k\) (\textit{Helpful}, \textit{Somewhat Helpful}, or \textit{Not Helpful}), with \(0\log 0=0\). We used a fixed denominator of \(\log 3\) for the three possible rating categories.

Metrics including rating counts, note age, text length, URL count, prior author note volume, and post-to-note delay were z-standardized. Prior Helpful mean, rating entropy, and author prior CRH share retained their original 0 to 1 scales, while binary indicators were coded as 0/1. No fixed effects were included, and standard errors were clustered by matching stratum. The coefficients represent adjusted associations with the log odds of observed population-sampled rating entry within the matched sample.

\textbf{Beyond the regression estimation, to further check whether the relationship between prior entropy and observed population-sampled rating entry varies across pre-event mean \textit{Helpful} score, we estimated a logistic regression including prior rating entropy, mean \textit{Helpful} score, and their interaction.} In this additional analysis, prior entropy was measured using the same normalized rating entropy as in the main model. We treated mean Helpful score as a continuous variable and included rating entropy, mean Helpful score, and their interaction:

\begin{equation}
\begin{aligned}
\operatorname{logit}\!\left[
\Pr(\mathrm{POPEntry}_i=1\mid D_i,H_i,\mathbf{X}_i)
\right]
={}& \alpha+\beta_1D_i+\beta_2H_i \\
&+\beta_3(D_i\times H_i)
+\boldsymbol{\gamma}^{\top}\mathbf{X}_i.
\end{aligned}
\label{eq:pop_disagreement_interaction}
\end{equation}

Here, $\mathrm{POPEntry}_i$ indicates whether note $i$ received a population-sampled rating, and $\operatorname{logit}(p)=\log[p/(1-p)]$ converts a probability to log odds. $D_i$ denotes the note's normalized pre-event rating entropy on its original 0--1 scale, and $H_i$ denotes its pre-event mean Helpful score centered at the estimation-sample mean. $\alpha$ is the intercept, $\beta_1$ and $\beta_2$ are the coefficients of rating entropy and centered Helpful score, respectively, and $\beta_3$ is the interaction coefficient. The interaction term $D_i\times H_i$ models the association between prior entropy and observed population-sampled rating entry, which varies across the continuous \textit{Helpful} score range. $\mathbf{X}_i$ contains the remaining controls and fixed effects, and $\boldsymbol{\gamma}$ is their coefficient vector.

For the entropy-by-\textit{Helpful} interaction model, we controlled for prior \textit{DEFAULT} rating volume, recent rating activity and concentration, note age, text length, URL count, author-selected reason counts, note format, and the trustworthy sources flag. Rating counts, note age, text length, and URL count were transformed using $\log(1+x)$. We included fixed effects for the calendar month and the note's age group at the event. Calendar month was the month when the event occurred, and the note's age was measured as the time elapsed from note creation to the event. These fixed effects allowed baseline POP-entry probabilities to differ across calendar months and note age groups. Standard errors were clustered by matching stratum.

Across levels of mean \textit{Helpful} score, we estimated the average predicted probability difference associated with a 0.1-unit increase in rating entropy, retaining the other covariates at their observed values. We report these contrasts in percentage points with pointwise 95\% CI as the intervals are intended to quantify uncertainty at each specified \textit{Helpful} score value~\cite{gsteiger2011simultaneous}; the entropy interaction coefficient provides the overall test of whether the association varies with the \textit{Helpful} score. We used the delta method to calculate the 95\% CI because the probability differences were derived from the fitted logistic regression. This method accounts for uncertainty in the estimated coefficients and the clustering of observations within matching strata, and is commonly used for marginal effects in non-linear models~\cite{oehlert1992note,mize2019best}.
\subsubsection{Sampling}

We compared notes that received a population-sampled rating with notes that received none during the observation period. For each treated note, the event was its first observed population-sampled rating. We selected a control note created in the same calendar month and assigned it a pseudo-event at the same note age, ensuring the time between receiving first POP rating and the creation of a note is same as the time between pseudo-event and the creation of a note. All covariates were measured before the actual or pseudo-event. The entropy analysis further required at least five prior \textit{DEFAULT} ratings to avoid estimating entropy caused by extremely sparse category counts~\cite{paninski2003estimation} and to be consistent with the platform's requirement that a note receive at least five total ratings before it can become eligible to leave NMR.


\subsection{Effects on Rating (RQ1)}

We next examined the effects of population-sampled ratings on different note rating aspects, including note resolution status and resolution speed. 

\subsubsection{Metrics}

\textbf{\textit{CRH and Non-NMR.}} We examined whether notes reached CRH or left NMR after the aligned event. Non-NMR denotes either CRH or CRNH. We summarized cumulative attainment over the follow-up period.

\textbf{\textit{Speed.}} We examined the elapsed time from the aligned event to the first subsequent CRH or non-NMR status. We measured rating accumulation using the time to the first 20 ratings and the seven-day rating rate. The seven-day window covers a complete weekly cycle, reducing differences caused by weekday and weekend activity patterns. We used 20 ratings as an interpretable accumulation milestone because prior research found that approximately 20 user ratings were sufficient to produce stable rankings in an online innovation community~\cite{riedl2013effect}. 

\textbf{\textit{Helpfulness difference.}} For each note with both a default rating and a population-sampled rating, we computed the helpfulness difference as $\Delta_i = \text{mean}_{i, \text{POP}} - \text{mean}_{i, \text{DEFAULT}}$, where $\text{mean}_{i, \text{POP}}$ and $\text{mean}_{i, \text{DEFAULT}}$ represent the mean helpfulness scores from the population-sampled and default ratings, respectively. This reflected the rating tendency differences between population-sampled rating and default rating.

\textbf{\textit{Helpfulness lift.}} For each note with both DEFAULT and population-sampled ratings, we defined lift as the mean helpfulness score across all DEFAULT and population-sampled ratings minus the mean score based on DEFAULT ratings alone, which measures the change in the mean helpfulness score associated with including population-sampled ratings.

\subsubsection{Estimation Model}

\textbf{\textit{Helpfulness regression.}} We estimated the following model for the helpfulness difference $\Delta_i$, including author fixed effects:

\begin{equation}
\begin{aligned}
\Delta_i
&= \alpha_{a(i)}
+ \beta_{\text{mis}} M_i \\
&\quad + \beta_{\text{other}} (MO_i)
+ \beta_{\text{fact}} (MF_i)
+ \beta_{\text{manip}} (MN_i)
+ \beta_{\text{outdated}} (MOU_i) \\
&\quad + \beta_{\text{context}} (MC_i)
+ \beta_{\text{unverified}} (MU_i)
+ \beta_{\text{satire}} (MS_i) \\
&\quad + \beta_{\text{CRH}} H_i
+ \beta_{\text{CRNH}} N_i
+ \beta_{\text{media}} \text{Media}_i
+ \beta_{\text{collab}} \text{Collab}_i \\
&\quad + \beta_{\log n}\log(1+\text{pop\_n}_i)
+ \beta_{\text{age}} \text{AgeAtFirstPOP}_i
+ \beta_{\text{after}} \text{AfterDef}_i
+ \varepsilon_i
\end{aligned}
\end{equation}

where $\alpha_{a(i)}$ captures the fixed effect for the author of note $i$, and $\varepsilon_i$ is the error term. We selected independent variables across post-level, note-level, and rating-level characteristics.

\textit{Post-level characteristics.}
Prior research on Birdwatch has examined misleading classifications and author-selected reasons~\cite{prollochs2022community}, motivating our examination of whether the helpfulness difference varies across these categories. $M_i$ indicates whether the note author classified the corresponding post as potentially misleading. Conditional on being a misleading note, a series of binary indicators capture specific sub-reasons: other reasons ($MO_i$), factual errors ($MF_i$), manipulated media ($MN_i$), outdated information ($MOU_i$), missing context ($MC_i$), unverified claims stated as facts ($MU_i$), and satire ($MS_i$).

\textit{Note-level characteristics.}
Prior research has examined note status stability and rating behaviour around note display~\cite{chuai2026consensus}, motivating our consideration of note status and lifecycle stage. $H_i$ and $N_i$ indicate whether the note's current status is CRH or CRNH, respectively. We additionally considered note format: $\text{Media}_i$ and $\text{Collab}_i$ indicate whether the note is a media note or a collaborative note, respectively. $\text{AgeAtFirstPOP}_i$ measures the days elapsed from note creation to its first population-sampled rating.

\textit{Rating-level characteristics.}
Prior research has examined changes in rating volume and rating leaning over time~\cite{chuai2026consensus}, motivating our consideration of rating volume and timing. To account for these characteristics, $\log(1+\text{pop\_n}_i)$ represents the log-transformed count of population-sampled ratings~\cite{eckles2021bias}. $\text{AfterDef}_i$ is a dummy variable indicating whether the timestamp of the first population-sampled rating postdates that of the first DEFAULT rating.

\textbf{\textit{Different effects across note groups.}} 
Since the association between baseline helpfulness and lift may be non-linear, we included both the linear and quadratic terms of the note-level mean DEFAULT helpfulness score. We modelled lift as a continuous function of this score. This quadratic specification, which has also been used to examine nonlinear relationships in online community research~\cite{shen2018person}, allows the relationship to vary with baseline helpfulness without imposing a strictly linear functional form. We report heteroskedasticity-robust pointwise 95\% CI and use Spearman's rank correlation to summarize the overall rank association.
\subsubsection{Sampling}

For the note-level analysis in RQ1, the treatment event was defined as the timestamp of each note's first observed population-sampled rating. Treated notes were restricted to those that still remained NMR at the treatment event, and matched control notes were restricted to have no population-sampled rating and were assigned a pseudo-treatment event in the same month and at a similar note age. 

We used PSM to construct note pairs and compared the two groups. The pre-event matching variables included DEFAULT rating count, mean helpfulness, and the standard deviation of helpfulness scores.

\subsection{Effects on Raters (RQ2)}
\label{sec:methods_raters}

We examined raters' rating behaviour changes after they addressed their first population-sampled rating, and how these changes varied across rater characteristics.

\subsubsection{Metrics}

\textbf{\textit{Rater behaviour metrics.}}
Prior research on crowd task notifications has examined both whether workers participate and how much they contribute~\cite{bashirieh2017nudge,kim2016studying}. Research on Community Notes has further considered rating volume and direction, the timing and selection of ratings, and the reasons provided for ratings~\cite{pilarski2024community,prollochs2022community,chuai2026consensus}. Building on these dimensions, we selected eight metrics covering participation intensity, rating selection, evaluative tendency, and diagnostic expression.

\textit{Not Helpful share} and \textit{mean rating score} summarize negative evaluations and average evaluative tendency, while \textit{diagnostic-tag-use intensity} measures the extent of reason-tag selection rather than evaluation quality.

\textit{Participation intensity} captures both whether raters participate and the number of ratings. It includes \textit{Ratings per active day}, defined as the mean number of \textit{DEFAULT} ratings submitted on days when a rater was active, and \textit{Daily active probability}, defined as the proportion of observed days on which the rater submitted at least one \textit{DEFAULT} rating.

\textit{Rating selection} captures the timing of ratings and the types of notes selected for evaluation. It includes \textit{Inter-rating interval}, defined as the mean elapsed time between consecutive DEFAULT ratings; \textit{Age of rated notes}, defined as the mean elapsed time between note creation and rating submission; and \textit{Misleading Note share}, defined as the proportion of \textit{DEFAULT} ratings assigned to notes whose authors classified the corresponding posts as misleading.

\textit{Evaluative tendency} captures the direction of raters' subsequent evaluations. It includes \textit{Not Helpful share}, the proportion of DEFAULT ratings marked as \textsc{Not Helpful}, and \textit{Mean rating score}, calculated on a 0 to 1 scale using the platform's official rating-to-score mapping.

Finally, \textit{Diagnostic expression} captures how extensively raters provide reasons for their evaluations. It is measured using \textit{Daily diagnostic-tag-use intensity}, defined as the mean number of diagnostic reason tags selected per rating on each active day. These tags provide structured explanations for ratings and are also used by the platform when determining and presenting note statuses.

\subsubsection{Estimation Model}

\textbf{\textit{DID analysis.}}
For each metric, we estimated a DID model comparing the change from the seven pre-event days to the seven post-event days among POP-exposed raters with matched control raters. The treatment event day was excluded to avoid mixing pre- and post-event behaviour within the same day and including the mechanical contribution of the index rating itself. The resulting p-values were adjusted across the eight outcomes using the Benjamini--Hochberg procedure~\cite{benjamini1995controlling}.

\textbf{\textit{Effect sizes and statistical precision.}}
We report each DID estimate and its 95\% confidence interval in the outcome's original unit. To assess sensitivity to smaller effects, we also calculated the approximate minimum detectable effect (MDE) at 80\% power and a two-sided significance level of $\alpha=.05$:
\begin{equation}
\mathrm{MDE}
=
\left(z_{.975}+z_{.80}\right)
\times \mathrm{SE}(\widehat{\mathrm{DID}}),
\end{equation}

where \(z_{.975}\) and \(z_{.80}\) are standard normal quantiles, and \(\mathrm{SE}(\widehat{\mathrm{DID}})\) is the estimated standard error for the corresponding outcome. These MDEs describe approximate sensitivity under the observed standard errors and use an unadjusted significance threshold. 

\textbf{\textit{Heterogeneity analysis.}}
Prior research has examined contributor activity, rating direction, and rating dynamics in Community Notes~\cite{prollochs2022community,chuai2026consensus}. We also drew on the platform's definitions of contributor standing, rating scores, and diagnostic reason tags. In particular, we referred to official metrics and used \textit{Earned in} to capture contributors' platform standing and \textit{Expansion+} to capture their modeling context. Based on these conceptual and platform-specific dimensions, we grouped the predictors into contributor standing, pre-event behaviour, and characteristics of the first POP rating.

We further conducted regression analyses to explore which rater characteristics were associated with changes in the eight behavioural outcomes. We estimated a separate regression for each outcome:

\[
Y_i^{(k)}
=
\alpha_k
+
\mathbf{S}_i^{\top}\boldsymbol{\beta}_k^{(S)}
+
\mathbf{T}_i^{\top}\boldsymbol{\beta}_k^{(T)}
+
\mathbf{B}_i^{\top}\boldsymbol{\beta}_k^{(B)}
+
\mathbf{F}_i^{\top}\boldsymbol{\beta}_k^{(F)}
+
\varepsilon_i^{(k)},
\qquad
k\in\{1,\ldots,8\}.
\]

Here, \(Y_i^{(k)}\) represents the matched pair DID change in outcome \(k\)~\cite{stuart2014using}. The vector \(\mathbf{S}_i\) captures contributor status based on Community Notes user enrollment, \(\mathbf{T}_i\) captures contributor tier indicators, \(\mathbf{B}_i\) captures pre-event behaviour, and \(\mathbf{F}_i\) captures characteristics of the first population-sampled rating. All eight predictors were included simultaneously in each outcome model. Specifically, the model includes:
\begin{description}
    \item[Contributor standing (\(\mathbf{S}\), \(\mathbf{T}\)):]
    \textit{Earned in} indicates whether the rater was in the platform's earned-in state at the treatment event, having accumulated sufficient Rating Impact to be eligible to write notes.
    \textit{Expansion+} indicates whether the rater had been assigned to the ExpansionPlus modeling population at the treatment event.

    \item[Pre-event behaviour (\(\mathbf{B}\)):]
    Prior research has examined rating volume and rating leaning across contributor groups~\cite{chuai2026consensus}, motivating our consideration of pre-exposure behaviour.
    \textit{Baseline helpfulness}, the mean helpfulness score across the rater's pre-event DEFAULT ratings using the platform's rating score mapping;
    \textit{Baseline ratings per active day}, the number of pre-event DEFAULT ratings divided by the number of pre-event days on which the rater submitted at least one DEFAULT rating;
    \textit{Fresh-note share}, the proportion of pre-event ratings assigned to notes created within the 24 hours preceding each rating; and
    \textit{Baseline diagnostic-tag intensity}, the mean number of diagnostic reason tags per pre-event DEFAULT rating.

    \item[First POP rating characteristics (\(\mathbf{F}\)):]
    Prior research has examined rating dynamics around note display~\cite{chuai2026consensus}. We additionally considered the score and note age at the first POP rating to characterize the initial rating context.
    \textit{First POP rating score}, calculated using the official score-mapping method, and
    \textit{First POP note age}, the elapsed time between the creation of a note and the rater's first POP rating on that note.
\end{description}

Continuous regressors and DID outcomes were standardized to z-scores within each estimation sample~\cite{schielzeth2010simple}, while binary indicators were coded as 0/1. The coefficients of continuous predictors therefore represent the change in each outcome, measured in standard deviations, associated with a one-standard-deviation increase in the corresponding predictor. The coefficients of Earned in and Expansion+ represent differences in the standardized outcome between raters in the corresponding category and those outside it. Models were fitted using OLS with HC1 heteroskedasticity-robust standard errors \cite{white1980heteroskedasticity,mackinnon1985some}. No interaction terms or fixed effects were included. The resulting p-values were adjusted across the 64 reported coefficients using the Benjamini--Hochberg procedure \cite{benjamini1995controlling}. 


We also conducted two one-sided equivalence tests using standardized bounds of \(\pm0.05\), \(\pm0.10\), and \(\pm0.20\) standard deviations \cite{schuirmann1987comparison,lakens2017equivalence}. Because these bounds were not defined a priori as substantively meaningful minimum effect sizes, we report them as sensitivity analyses. 

\subsubsection{Sampling}

\textbf{\textit{Population-sampled groups and matched control groups.}}
For the rater-level analysis in RQ2, the treatment event was defined as the timestamp when a rater submitted their first population-sampled rating, to align comparisons around the first observed POP rating. The matched control group included raters who never submitted a population-sampled rating but submitted a DEFAULT rating on the same note within two hours of the treated rater's treatment event. Note content, note status, and the stage of the note's rating lifecycle were also considered. Treated and control raters were further matched by their pre-treatment rating volume, active days, weekly participation trajectories, mean helpfulness, Helpful share, tag-use patterns, and length of rating experience. The timestamp of the corresponding control rating was assigned as the pseudo-treatment event.

We conducted three additional analyses on the assumptions and sensitivity of the rater-level DID comparisons. First, we tested the differences between treated and control groups across pre-event days to examine whether these groups had similar pre-event metrics. Second, we re-estimated the DID results by excluding the index notes from the matched pairs, and subsequently by adopting a more conservative approach that excluded ratings for any note that had been included in the population sampling process. These analyses aimed to assess whether the estimates were influenced by ratings directly associated with the POP process. Third, we re-estimated all eight contrasts using the full archived matched-rater pool to assess sensitivity to sample construction. All matched samples passed such tests, and are non-significant for such comparisons between treatment and control groups.


\section{RQ1: Observed Population-Sampled Rating Entry and Subsequent Note Outcomes}

We first examine which notes are likely to receive population-sampled ratings by examining post-, note-, and rating-level factors. We next investigate whether the association between disagreement and observed population-sampled entry varies with prior Helpful support.

\subsection{Correlations of Contextual Factors on Population-Sampled Rating}

\begin{figure}
\centering
\includegraphics[width=\textwidth]{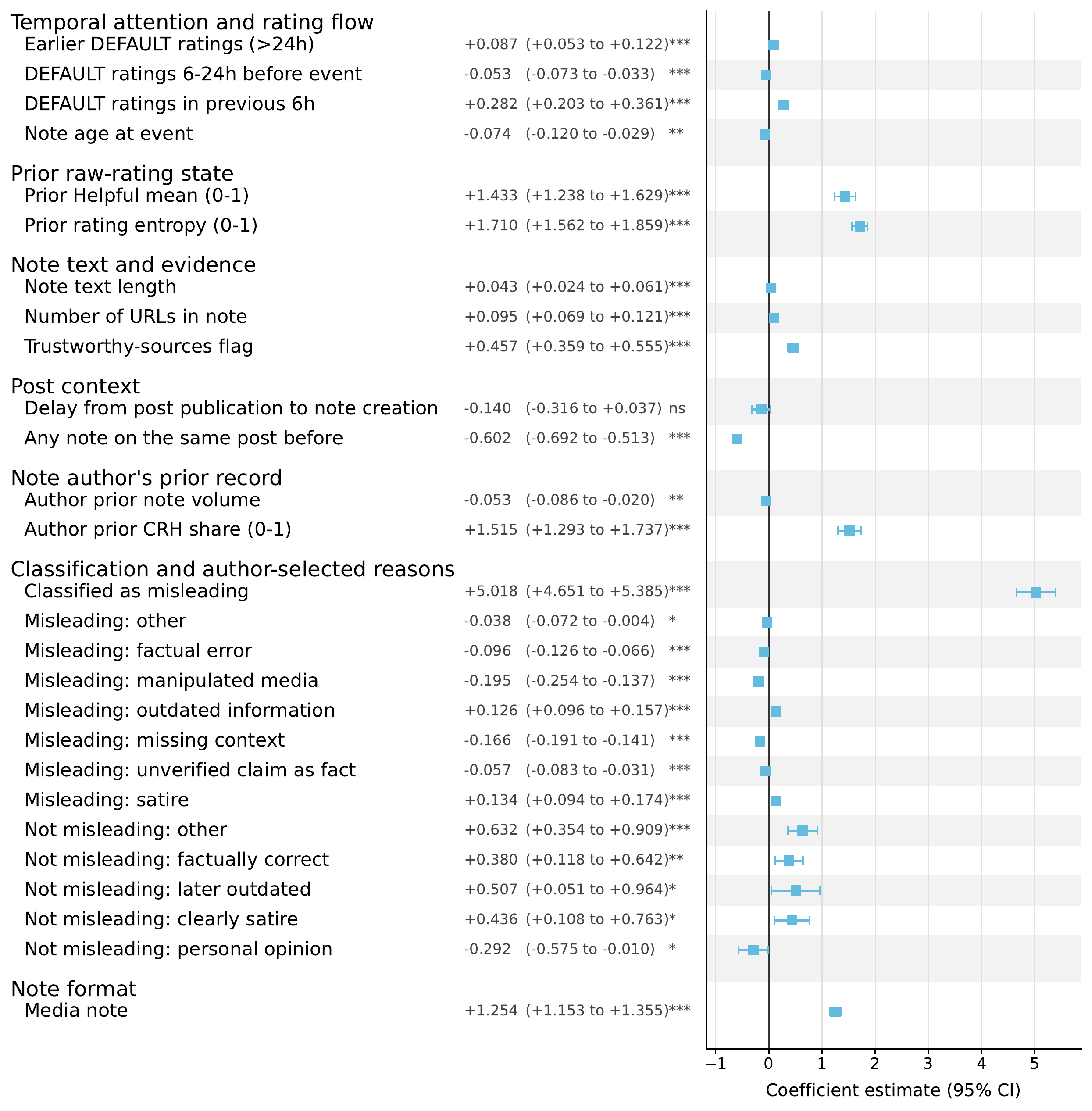}
\caption{Correlations of whether a note received population-sampled rating across different factors. Error bars denote 95\% confidence intervals.}
\Description{A horizontal forest plot of regression coefficients for observed population-sampled rating entry, grouped by rating activity, prior rating state, note text and evidence, post context, author history, classification and author-selected reasons, and note format. Squares mark estimates, horizontal bars show 95\% Confidence Intervals, and a vertical line marks zero. Recent rating activity, prior Helpful support, rating entropy, media-note format, and the author's prior CRH share are positively associated with entry. Note age and the presence of an earlier note on the same post are negatively associated with entry.}
\label{fig:factors}
\end{figure}

As shown in Figure~\ref{fig:factors}, we examine which pre-event characteristics are associated with whether a note receives a population-sampled rating.

\textbf{Post-level characteristics.}

\textbf{\textit{Post context.} Prior note presence on the same post reduces the likelihood of population-sampled entry.} Specifically, notes written for posts that already have existing notes (Any note on the same post before) are less likely to receive population-sampled ratings ($\beta=-0.602$, 95\% CI $[-0.692,-0.513]$, $q<.001$). In contrast, the time elapsed between post publication and note creation shows no significant association.

\textbf{\textit{Classifications and author-selected reasons.} Author-assigned classifications and specific rationales have varied effects on population-sampled entry.} These classifications describe the authors' assessments of the corresponding posts rather than independently verified post accuracy. Notes classified as addressing misleading content (Classified as misleading) show a strong positive association with an observed population-sampled rating ($\beta=5.018$, 95\% CI $[4.651,5.385]$, $q<.001$), a result tied to the accompanying reason indicators. Conditional on overall classification and other covariates, cited outdated information ($\beta=0.126$, 95\% CI $[0.096,0.157]$, $q<.001$) and satire ($\beta=0.134$, 95\% CI $[0.094,0.174]$, $q<.001$) are positively correlated with observed population-sampled entry. In contrast, factual error ($\beta=-0.096$, 95\% CI $[-0.126,-0.066]$, $q<.001$), manipulated media ($\beta=-0.195$, 95\% CI $[-0.254,-0.137]$, $q<.001$), missing context ($\beta=-0.166$, 95\% CI $[-0.191,-0.141]$, $q<.001$), and an unverified claim presented as fact ($\beta=-0.057$, 95\% CI $[-0.083,-0.031]$, $q<.001$) are negatively associated with selection.

For non-misleading notes, reasons such as factually correct ($\beta=0.380$, 95\% CI $[0.118,0.642]$, $q=.006$), later outdated ($\beta=0.507$, 95\% CI $[0.051,0.964]$, $q=.032$), clearly satire ($\beta=0.436$, 95\% CI $[0.108,0.763]$, $q=.011$), and other ($\beta=0.632$, 95\% CI $[0.354,0.909]$, $q<.001$) are positively associated with selection, while personal opinion is negatively associated ($\beta=-0.292$, 95\% CI $[-0.575,-0.010]$, $q=.044$).

\textbf{Note-level characteristics.}
\textbf{\textit{Note text and evidence.} Substantiated and comprehensive note content is positively correlated with population-sampled rating.} Note text length has a modest positive association ($\beta=0.043$, 95\% CI $[0.024,0.061]$, $q<.001$), as does the number of supporting URLs ($\beta=0.095$, 95\% CI $[0.069,0.121]$, $q<.001$). Notes flagged by authors as citing reliable sources are significantly more likely to receive population-sampled ratings ($\beta=0.457$, 95\% CI $[0.359,0.555]$, $q<.001$).

\textbf{\textit{Note format.} Media-rich notes are notably more likely to be selected than text-only notes.} Compared to non-media notes, media notes exhibit a pronounced positive association with receiving a population-sampled rating ($\beta=1.254$, 95\% CI $[1.153,1.355]$, $q<.001$).

\textbf{\textit{Note age.} Newer notes are prioritized for population-sampled evaluation.} Conditional on other covariates, note age is negatively associated with entry ($\beta=-0.074$, 95\% CI $[-0.120, -0.029]$, $q=.002$), indicating that older notes are less likely to be sampled.

\textbf{\textit{Note author's prior record.} Author reputation promotes sampling, whereas historical volume disincentivizes it.} Among authors with resolved notes, historical note volume is negatively correlated with receiving a population-sampled rating ($\beta=-0.053$, 95\% CI $[-0.086, -0.020]$, $q=.002$). Conversely, an author's historical CRH share (\textit{Author prior CRH share}) demonstrates a strong positive association ($\beta=1.515$, 95\% CI $[1.293, 1.737]$, $q<.001$), underscoring divergent effects between quality and quantity.

\textbf{Pre-event rating characteristics.}
\textbf{\textit{Temporal attention and rating flow.} Recent bursts and early inflows of ratings enhance entry likelihood, while intermediate activity decreases it.} The volume of DEFAULT ratings received within six hours prior to the event is positively associated with selection ($\beta=0.282$, 95\% CI $[0.203, 0.361]$, $q<.001$). Similarly, early ratings received over 24 hours prior exhibit a positive association ($\beta=0.087$, 95\% CI $[0.053, 0.122]$, $q<.001$). In contrast, ratings received during the intermediate 6--24-hour window have a negative association ($\beta=-0.053$, 95\% CI $[-0.073, -0.033]$, $q<.001$).

\textbf{\textit{Prior raw-rating state.} Higher consensus and viewpoint diversity in prior ratings both favour population-sampled rating.} Specifically, both the prior Helpful mean ($\beta=1.433$, 95\% CI $[1.238, 1.629]$, $q<.001$) and prior rating entropy ($\beta=1.710$, 95\% CI $[1.562, 1.859]$, $q<.001$) are strongly and positively associated with receiving a population-sampled rating.

\subsection{Effects of Disagreement and Strong Helpful Support}
\begin{figure}
    \centering
    \subfloat[Cumulative share of notes reaching CRH.]{
        \includegraphics[width=0.53\textwidth]
        {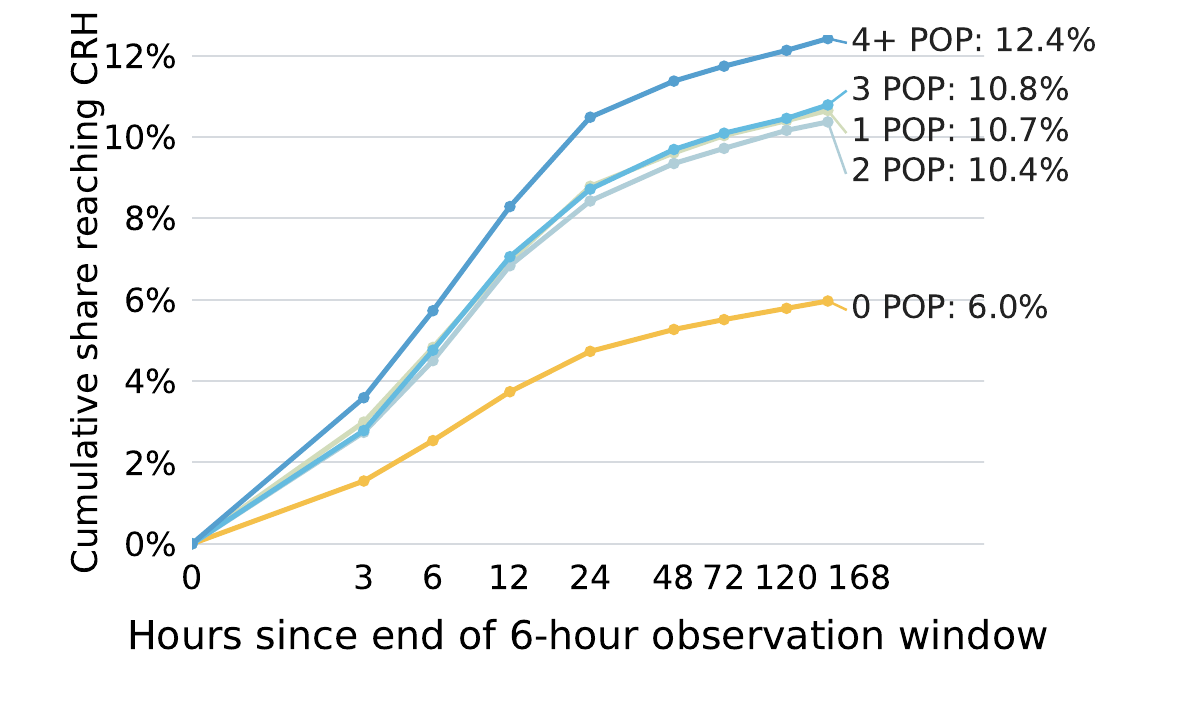}
        \label{fig:pop_crh_cumulative}
    }
    \subfloat[Difference in POP-entry probability.]{
        \includegraphics[width=0.47\textwidth]
        {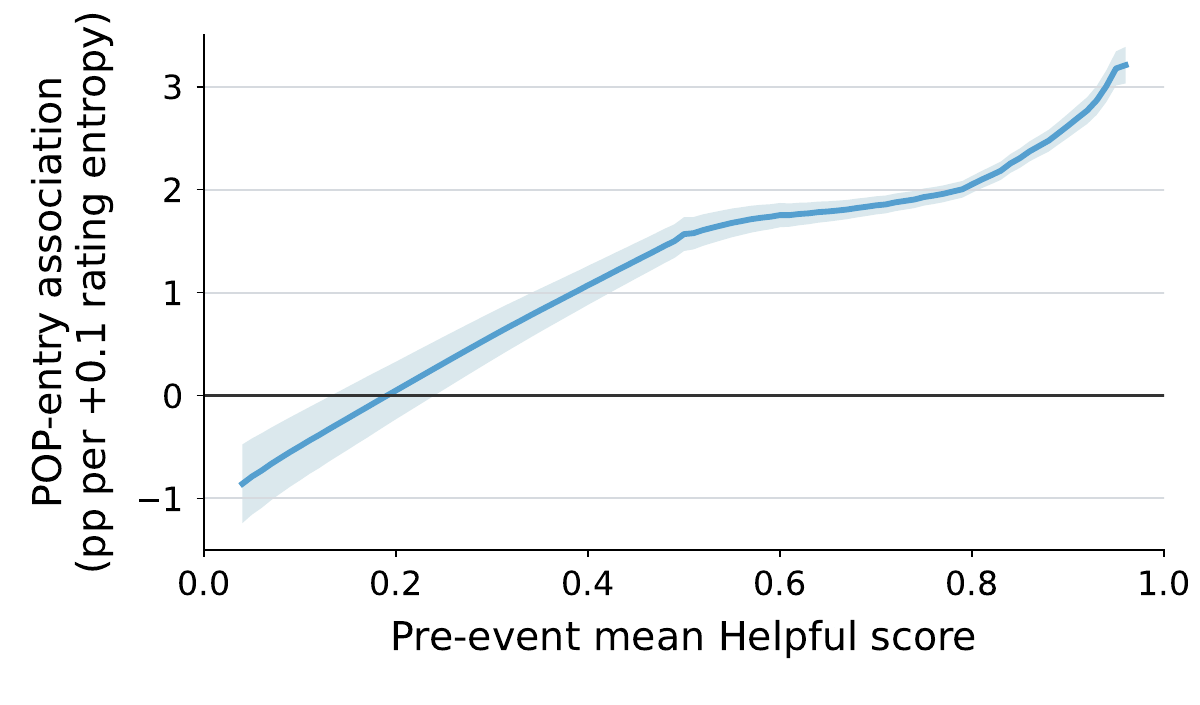}
        \label{fig:pop_entropy_interaction}
    }
    \caption{Observed POP entry and subsequent CRH attainment. 
    (a) Cumulative share of notes reaching CRH, grouped by the number of population-sampled ratings received during the six-hour observation window. The horizontal axis presents follow-up times from 0 to 168 hours. 
    (b) Percentage-point difference in POP-entry probability associated with a 0.1 increase in prior rating entropy across pre-event mean Helpful scores. Shading shows pointwise 95\% confidence intervals. Predictions are averaged over score-specific samples for which both entropy values are theoretically feasible.}
    \Description{Two graphs show note-level patterns related to population-sampled ratings. Graph (a) plots cumulative Currently Rated Helpful (CRH) attainment over 168 hours after a six-hour observation window, using non-linear time spacing. Notes with no POP ratings reach 6.0\% CRH attainment, compared with 10.4 to 10.8\% for one to three POP ratings and 12.4\% for four or more. Graph (b) plots the percentage-point difference in POP-entry probability associated with a 0.1 increase in prior rating entropy. This association rises with the pre-event mean Helpful score, changing from negative at low scores to positive at higher scores. Shading indicates the 95\% confidence interval.}
    \label{fig:pop_entry_and_crh}
\end{figure}

The above analysis showed that both prior Helpful mean and rating entropy were positively associated with observed POP entry. We further examined whether the relationship between prior rating entropy and observed POP entry varies depending on the pre-event mean Helpful score. As shown in Figure~\ref{fig:pop_entropy_interaction}, the adjusted probability contrasts were generally larger at higher pre-event mean Helpful scores. The Helpful-score-by-entropy interaction was positive and significant ($\beta=2.674$, 95\% CI $[2.410,2.938]$, $p<.001$).

At a mean Helpful score of 0.50, a 0.1-unit increase in prior rating entropy was associated with an estimated 1.57-percentage-point increase in POP-entry probability (95\% CI $[1.40,1.73]$). The corresponding estimate was 2.05 percentage points at a mean Helpful score of 0.80 (95\% CI $[1.97,2.13]$). At low Helpful levels, the association was negative or indistinguishable from zero. These contrasts were averaged over score-specific eligible samples.

These findings suggest that the association between rating entropy and observed POP entry depends on existing Helpful support. The results are consistent with observed POP entry being positively associated with residual rating dispersion among notes that have already received strong Helpful support.

\subsection{Associations with Subsequent Note Progression}

Having identified the characteristics associated with observed population-sampled rating entry, we next examined whether this entry was followed by differences in rating accumulation, resolution speed, resolution status, and rating patterns. These analyses provide evidence about whether observed population-sampled rating participation is aligned with the feature's stated goal of helping unresolved notes progress toward consensus.

As speed and quality are among the most important aspects of Community Notes~\cite{de2025supernotes,wojcik2022birdwatch}, we first examined whether the population-sampled rating on a specific note affect the note's speed and quality. We then examined whether this effect remains robust across post-, note-, and rating-level factors. 

\subsubsection{Effects on Note Resolution Speed}

We first explored how quickly notes reached a resolved status (Helpful or Not Helpful) after their first population-sampled rating. Among notes receiving population-sampled ratings that subsequently reached CRH, the median post-exposure time to CRH was 3.51 hours (M=15.44 hours), compared with 6.35 hours for matched controls (M=22.60 hours, Mann-Whitney $p < .001$). Among notes that receive population-sampled ratings which subsequently left NMR, the median post-exposure time to the first non-NMR status was 3.47 hours for treated notes (M=15.27 hours), compared with 6.08 hours for matched controls (M=21.54 hours, Mann-Whitney $p < .001$). These results show that notes with population-sampled ratings reach both CRH and non-NMR status significantly faster than controls. 

We found that population sampling mainly accelerated rating accumulation early in the note lifecycle. Notes with population-sampled ratings did not have a significantly different 7-day rating speed from matched controls (median 5.57 vs. 5.43 ratings per day, paired Wilcoxon $p=.926$), but notes with population-sampled ratings acquired their first 20 ratings significantly faster (median 1.74 vs. 2.63 hours, paired Wilcoxon $p < .001$). Therefore, population-sampled ratings seem to move notes toward a decision earlier in the cycle, instead of simply making raters rate more, as evidenced in Figure~\ref{fig:source_crh} and ~\ref{fig:source_non_nmr}.

\subsubsection{Effects on Note Resolution Status}

To explore whether the early number of population-sampled ratings was associated with notes' subsequent status, we grouped notes according to whether they received 0, 1, 2, 3, or more population-sampled ratings, and then calculated the cumulative proportion of notes that reach CRH from 0 to 168 hours after the observation window. Specifically, we focus on time points such as 12 hours, 24 hours, 7 days and 30 days, which captures the first half-day, the first full day, a complete weekly cycle, and a monthly cycle after the event.

As shown in Figure~\ref{fig:pop_crh_cumulative}, the cumulative CRH proportion increases most rapidly during the first 24 hours and then grows more slowly. Notes receiving at least one early POP rating consistently show higher cumulative CRH proportions than notes receiving none. By 168 hours, only 6.0\% of notes without an early POP rating reach CRH, compared with 10.4--12.4\% among notes receiving 1 to 4 POP ratings. Overall, the one-, two-, and three-POP groups showed similar cumulative CRH attainment, whereas notes receiving four or more POP ratings reached a somewhat higher cumulative CRH proportion.

\begin{figure}
    \subfloat[CRH.]{
        \includegraphics[width=0.5\textwidth]{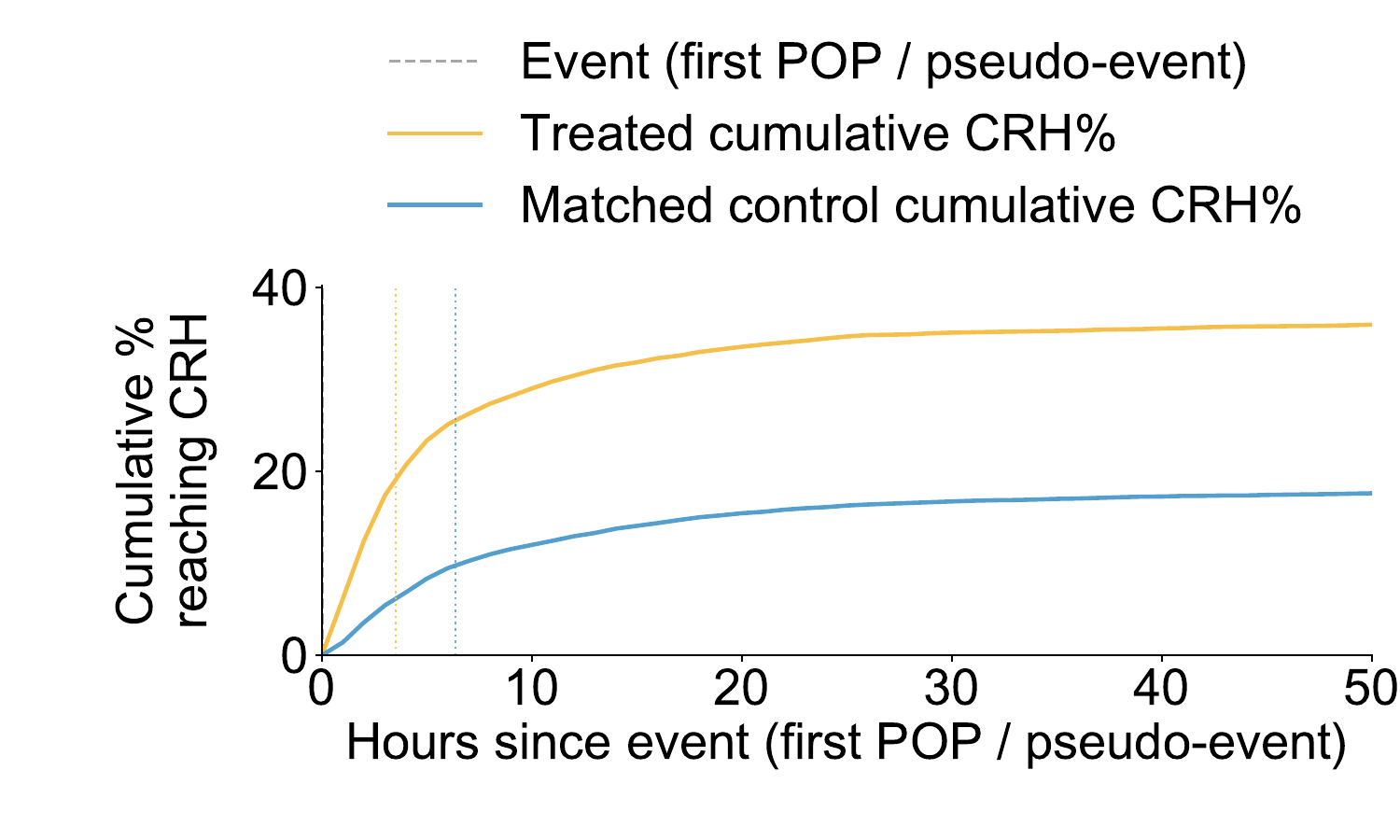}
        \label{fig:source_crh}
    }
    \subfloat[Non-NMR.]{
        \includegraphics[width=0.5\textwidth]{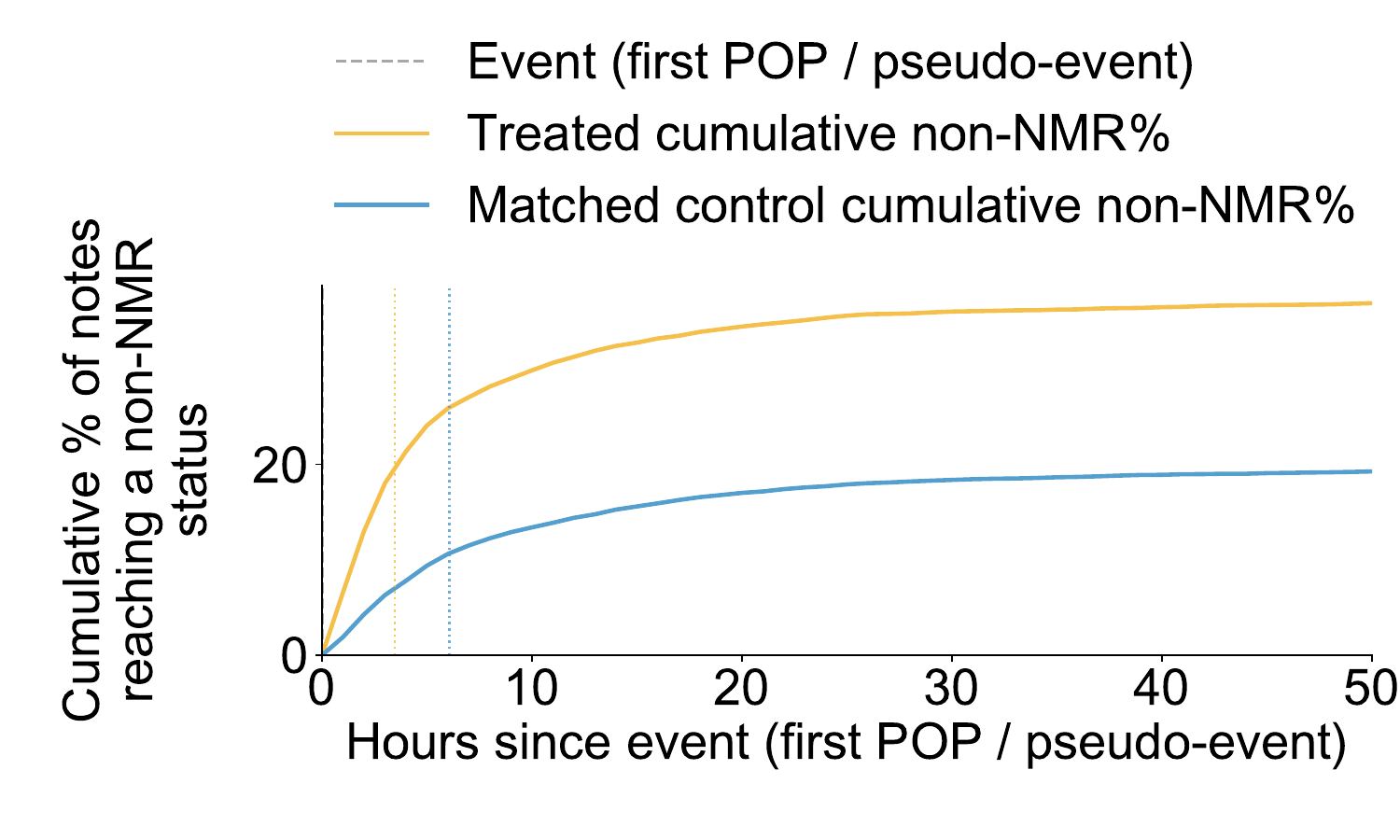}
        \label{fig:source_non_nmr}
    }
    \caption{The CRH and Non-NMR rate of notes receiving population-sampled ratings, compared to control groups that did not receive population-sampled ratings.}
    \Description{Two cumulative line graphs compare notes receiving population-sampled ratings with matched control notes over the first 50 hours after the first POP rating or corresponding control event. Graph (a) shows the percentage of notes reaching Currently Rated Helpful (CRH). Graph (b) shows the percentage of notes leaving Needs More Ratings (NMR) status for the first time. In both graphs, the orange curve for exposed notes rises more steeply and remains above the blue curve for matched controls. Most of the separation emerges during the first 12 hours.}
    \label{fig:source}
\end{figure}

Relative to matched controls, notes exposed to population-sampled ratings were 13.17 percentage points more likely to reach CRH within 12 hours (95\% CI $[12.70,13.65]$, $p<.001$) and 15.64 percentage points more likely within 24 hours (95\% CI $[15.11,16.16]$, $p<.001$).  The difference increased to 17.26 percentage points within seven days (95\% CI $[16.71,17.82]$, $p<.001$) and remained similar at 30 days (17.65 percentage points, 95\% CI $[17.09,18.21]$, $p<.001$). By 12 hours, 76.3\% of the seven-day difference had already emerged (bootstrap 95\% CI $[74.6,78.0]$); by 24 hours, this proportion reached 90.6\% ($[89.4,91.8]$). Among notes that reached CRH after the aligned event, the median lag was 3.51 hours for treated notes and 6.35 hours for matched controls. Overall, these results suggest that the separation in CRH attainment emerged primarily during the first day and persisted throughout the seven-day follow-up period, as shown in Figure~\ref{fig:source_crh}.

Compared with matched controls, the first non-NMR status arrived a median of 3.47 hours after exposure in treated notes versus 6.08 hours in matched controls (two-sided Mann--Whitney $p<.001$). The seven-day matched difference for leaving NMR was $+13.90$ percentage points (95\% CI $[13.27,14.46]$, $p<.001$), while the corresponding CRH difference was $+17.26$ percentage points (95\% CI $[16.71,17.82]$, $p<.001$). The implied difference in CRNH attainment was $-3.37$ percentage points (95\% CI $[-3.67,-3.08]$, $p<.001$). These results suggest that population-sampled ratings primarily help unresolved notes leave NMR by moving them toward CRH rather than CRNH, as shown in Figure~\ref{fig:source_non_nmr}.

\begin{figure}[!htbp]
    \includegraphics[width=\textwidth]{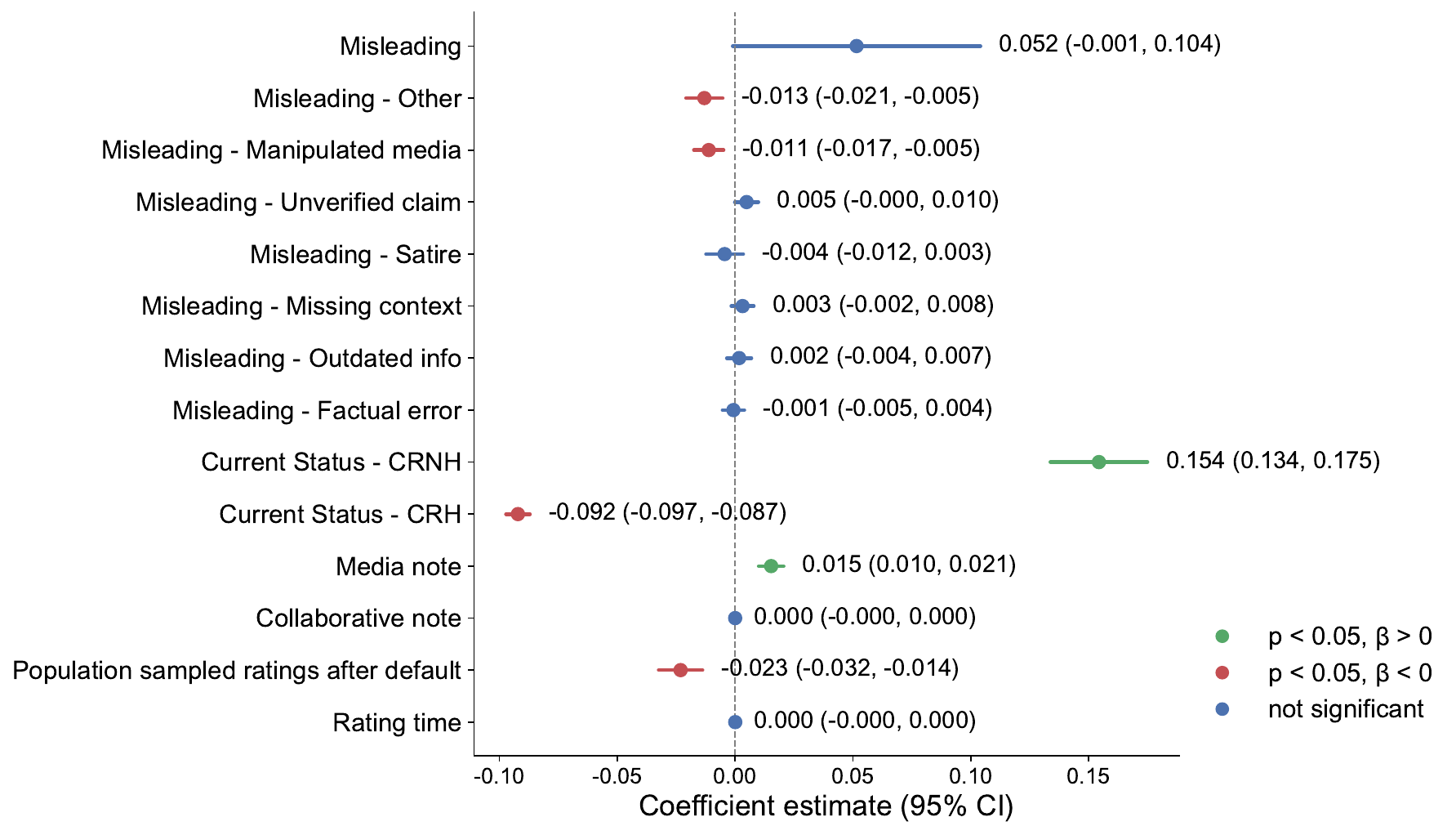}
    
    \caption{Regression results of note-related factors with respect to the mean raw helpfulness score difference between paired population-sampled and default ratings.}
    \Description{A horizontal forest plot shows associations between note-related factors and the mean helpfulness score difference between population-sampled and default ratings, calculated as POP minus DEFAULT. Points mark coefficient estimates, and bars show 95\% confidence intervals around a zero reference line. Green and red indicate significant positive and negative associations, respectively. Blue indicates non-significant estimates. Current note status shows the largest associations, that Currently Rated Not Helpful (CRNH) status is associated with a larger POP-minus-DEFAULT difference, whereas Currently Rated Helpful (CRH) status is associated with a smaller difference.}
    \label{fig:rating_source}
\end{figure}

\subsection{Correlation Analysis Around Subsequent Note Outcomes}

\textbf{Discrepancies between population-sampled and default ratings are strongly correlated with note content, existing status, and timing, showing that sampled ratings possess corrective potential.} We performed linear regression on the raw helpfulness rating difference comparing population-sampled rating and default rating. As shown in Figure~\ref{fig:rating_source}, we found that notes addressing misleading posts were marginally associated with a larger difference for population-sampled compared with default rating ($\beta=0.052$, $p = .05$). Specifically, notes with misleading reasons as \textit{Other} or \textit{Manipulated media} correlate with a smaller difference for population-sampled compared with default rating, while other categories are non-significant (Other: $\beta=-0.013$, $p < .001$, Manipulated media: $\beta=-0.011$, $p < .001$). 

Interestingly, notes that are CRNH significantly correlate with a larger difference between population-sampled rating and default ratings ($\beta=0.154$, $p < .001$), with those CRH correlating with a smaller difference between population-sampled rating and default ratings ($\beta=-0.092$, $p < .001$). This suggested that population-sampled ratings are shown to have the potential to change the status, as it goes opposite wide with default ratings.

Furthermore, we observed that media notes were positively associated with a larger difference between population-sampled rating and default rating ($\beta=0.015$, $p < .001$), which may be because multimedia content often elicits emotionally driven organic reactions from default raters, whereas the population-sampled ratings may be more rational. Receiving the first population-sampled rating after the first DEFAULT rating was associated with a smaller POP-minus-DEFAULT helpfulness difference ($\beta=-0.023$, $p<.001$), confirming the first-mover effect observed in prior work~\cite{gong2026characterizing}.

\begin{figure}
    \centering
    \includegraphics[
        width=0.75\textwidth,
        keepaspectratio
    ]{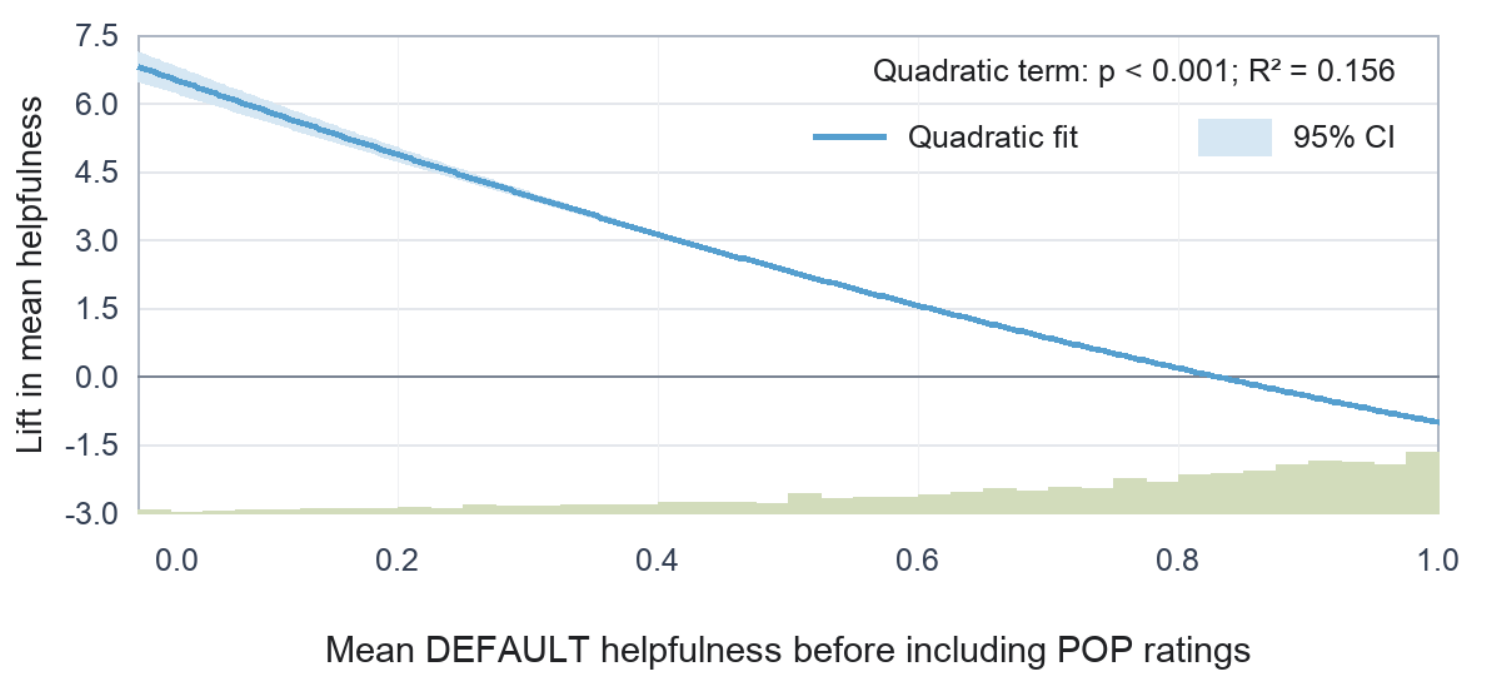}
    \caption{Continuous relationship between helpfulness and the lift associated with including population-sampled ratings. The pale green bars along the bottom show the distribution of notes across the mean DEFAULT helpfulness score. The solid line shows the quadratic regression estimate, and the blue shaded area shows the pointwise 95\% CI.}
    \Description{The figure shows how helpfulness lift varies with the mean default rating score for each note. Lift measures the change in mean helpfulness when population-sampled ratings are included. The fitted curve declines as default scores increase. Lift is positive at lower scores, crosses zero near 0.83, and becomes negative at higher scores. The blue line shows the quadratic fit. Shading indicates the 95\% confidence interval. Green bars show the distribution of default scores, with more notes at higher scores.}
    \label{fig:helpfulness_lift_continuous}
\end{figure}

\textbf{Trends across rating helpfulness.}
Figure~\ref{fig:helpfulness_lift_continuous} shows the fitted quadratic relationship between the note-level mean DEFAULT helpfulness score and lift.

The fitted quadratic relationship showed an overall downward pattern. Based on the same 186,405 notes, the fitted model was \(\widehat{\mathrm{Lift}}=6.811-10.156H+2.338H^2\), where \(H\) denotes the note-level mean DEFAULT helpfulness score. The quadratic term was statistically significant (\(p<.001\)), and the model had an \(R^2\) of .156. Mean lift tended to be lower as the note-level mean DEFAULT helpfulness score increased. Notes with lower DEFAULT helpfulness generally showed more positive lift after population-sampled ratings were included. The estimated lift approached zero as DEFAULT helpfulness increased and became negative near the upper end of the observed range. Consistent with this pattern, DEFAULT helpfulness was negatively correlated with lift (Spearman's $\rho=-0.436$, $p<.001$). These results suggest that the inclusion of population-sampled ratings was associated with more positive lift among notes with lower DEFAULT helpfulness and with smaller or negative lift among notes with higher DEFAULT helpfulness.

\begin{figure*}
    \centering
    \includegraphics[
    width=\textwidth
    ]{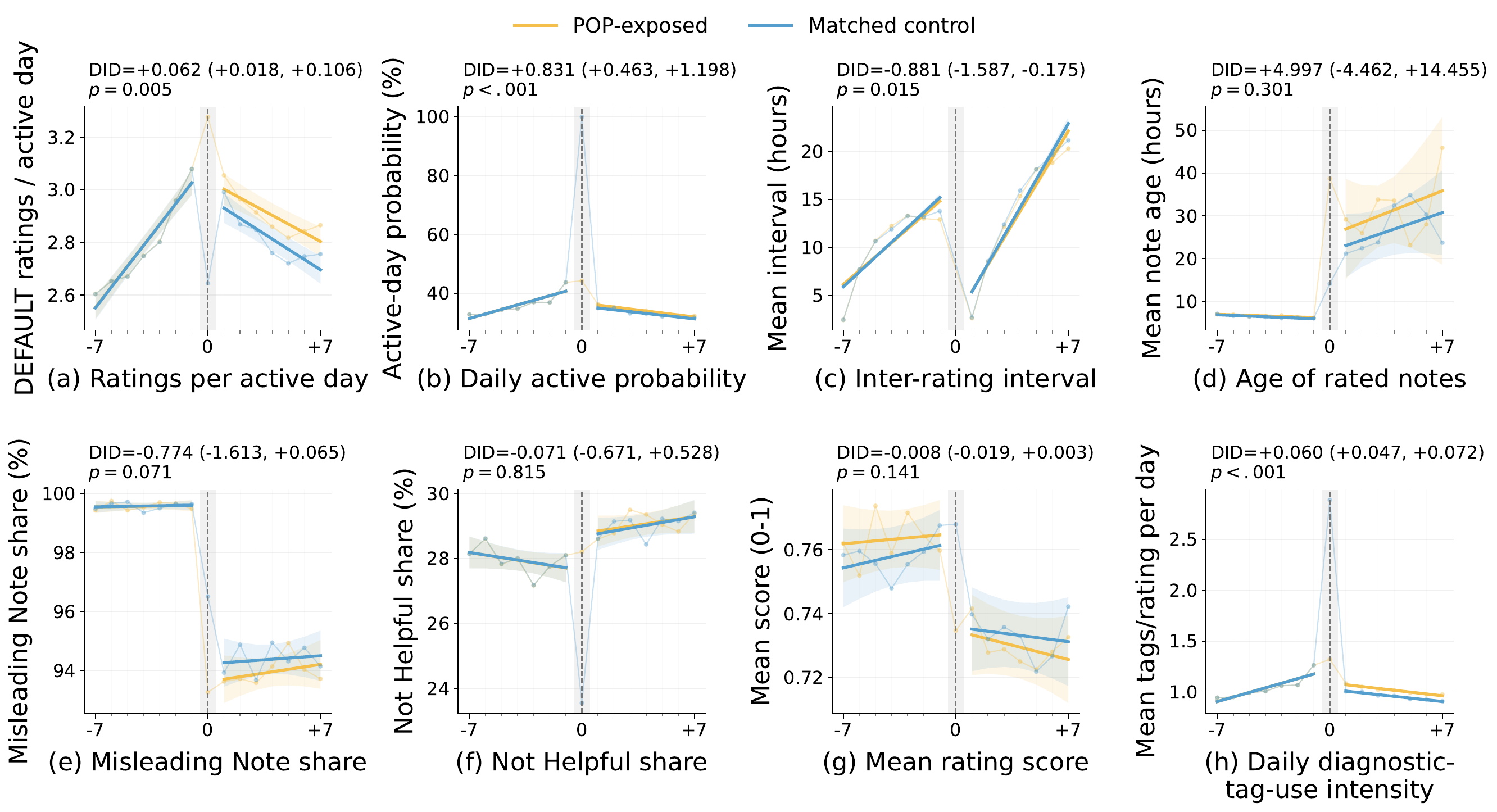}
    \caption{Behaviour of POP-exposed and matched control raters during the seven days before and after the event. (a) Mean number of DEFAULT ratings per active day. (b) Percentage of raters submitting at least one DEFAULT rating each day. (c) Mean interval between consecutive DEFAULT ratings, in hours. (d) Mean age of notes when rated, in hours. (e) Percentage of DEFAULT ratings on notes addressing posts classified as misleading by note authors. (f) Percentage of DEFAULT ratings marked Not Helpful. (g) Mean DEFAULT helpfulness score on a scale from 0 to 1. (h) Mean number of diagnostic tags per DEFAULT rating on active days. Orange and blue lines represent POP-exposed raters and matched controls, respectively. Shaded areas show 95\% confidence intervals. Day zero marks the first observed POP rating or the corresponding control event and is excluded from estimation.}
    \Description{Eight graphs compare the behaviour of raters who completed a population-sampled rating with that of matched controls. Each diagram covers seven days before and after the first observed POP rating or the corresponding control event. The outcomes include rating activity, rating pace, note selection, rating scores, and diagnostic tag use. Orange lines represent POP raters, and blue lines represent controls. Shading shows 95\% confidence intervals. A vertical dashed line marks day zero, which is excluded from estimation. Relative to controls, POP raters show increases in ratings per active day, daily active probability, and diagnostic tag use, along with shorter intervals between ratings. The other four outcomes show no significant changes.}
    \label{fig:rater_cumulative_did}
\end{figure*}

\section{RQ2: Rater Behaviour Around the First Population-Sampled Rating}

We wanted to understand whether raters' behaviour changes around their first observed population-sampled rating, including their participation, rating cadence and note selection, evaluative tendency, and diagnostic-tag use. We were also interested in whether these changes vary across rater characteristics and the context of their first population-sampled rating.

\subsection{Effects of Population-Sampled Rating}

We examined whether POP-exposed raters have different behavioural changes from matched control raters during the seven days before and after the treatment event. The eight behavioural outcomes and the estimation procedure are described in Section~\ref{sec:methods_raters}.

\textbf{Overall, four metrics showed significant changes across the 7-day period: \textit{ratings per active day}, \textit{daily active probability}, \textit{inter-rating interval}, and \textit{daily diagnostic-tag-use intensity}.} We found no significant changes in the other four metrics within the 7-day post-event period.

\textbf{As shown in Figure~\ref{fig:rater_cumulative_did}, we found significant increases in both \textit{ratings per active day} and \textit{daily active probability} around the first POP rating across 7 days.} The former increased by 0.062 relative to the matched controls (95\% CI $[0.018,0.106]$, $p=.005$, $q=.015$), while the latter increased by 0.831 percentage points (95\% CI $[0.463,1.198]$, $p<.001$, $q<.001$).

\textbf{For rating pace and selection, \textit{inter-rating interval} decreased, while neither \textit{age of rated notes} nor \textit{misleading note share} showed a significant change.} The \textit{inter-rating interval} decreased by 0.881 hours relative to the matched controls (95\% CI $[-1.587,-0.175]$, $p=.015$, $q=.029$), remaining significant after correction. Neither \textit{age of rated notes} ($\mathrm{DID}=+4.997$ hours, 95\% CI $[-4.462,14.455]$, $p=.301$, $q=.344$) nor \textit{misleading note share} ($\mathrm{DID}=-0.774$ percentage points, 95\% CI $[-1.613,0.065]$, $p=.071$, $q=.113$) showed a significant change.

\textbf{Regarding evaluative tendency, neither \textit{Not Helpful share} nor \textit{Mean rating score} showed a significant change.} The corresponding estimates were $-0.071$ percentage points for \textit{Not Helpful share} (95\% CI $[-0.671,0.528]$, $p=.815$, $q=.815$) and $-0.008$ for \textit{Mean rating score} (95\% CI $[-0.019,0.003]$, $p=.141$, $q=.187$).

\textbf{\textit{Daily diagnostic-tag-use intensity} also increased and remained significant after correction.} It increased by $0.060$ in the daily mean number of tags per rating relative to the matched controls (95\% CI $[0.047,0.072]$, $p<.001$, $q<.001$). This finding indicates greater subsequent diagnostic-tag use and may reflect increased familiarity with the platform's diagnostic criteria.


\subsection{Correlation Analysis}

While population-sampled ratings have limited effects on raters overall, these effects may vary across heterogeneous raters. We therefore evaluated whether behavioural changes differ across specific rater types that are defined by \X.

\begin{figure*}
    \centering
    \includegraphics[
        width=\textwidth
    ]{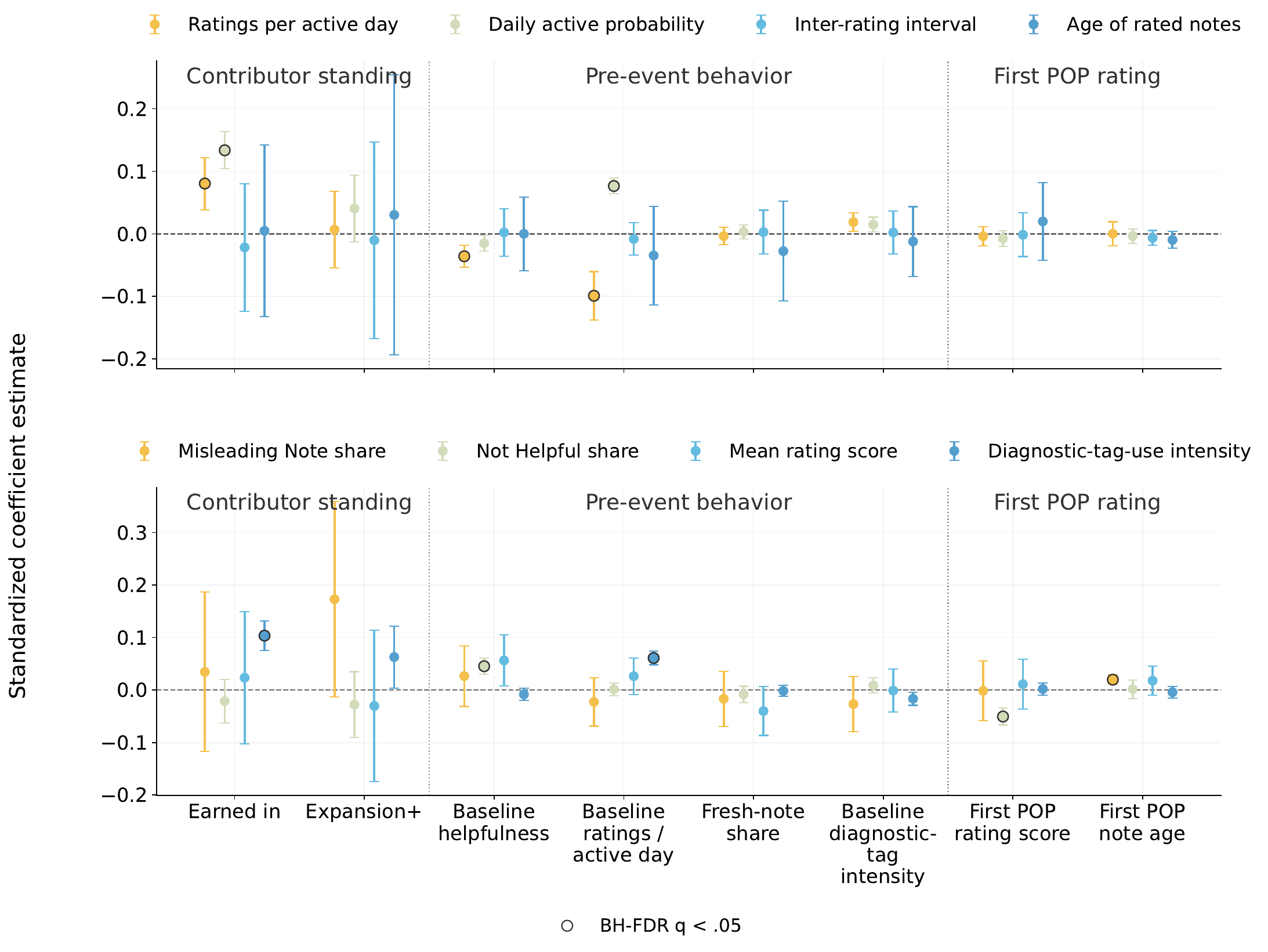}
    \caption{Heterogeneity in behavioural changes following raters' first observed population-sampled rating. The two panels report standardized coefficients from separate linear regression models for eight behavioural outcomes. Points show coefficient estimates, and error bars show 95\% confidence intervals. Black-outlined points remain significant after Benjamini-Hochberg correction.}
    \Description{Two vertically stacked coefficient plots relate contributor standing, pre-event behaviour, and first POP rating characteristics to matched pair-level behavioural changes. The upper diagram covers rating activity, inter-rating interval, and age of rated notes; the lower diagram covers misleading note share, Not Helpful share, mean rating score, and diagnostic-tag-use intensity. Colors distinguish outcomes, points show standardized coefficients, and bars show 95\% confidence intervals. Black outlines identify ten associations that remain significant after Benjamini--Hochberg correction. Earned-in status is positively associated with changes in ratings per active day, daily active probability, and diagnostic tag using intensity.}
    \label{fig:rater_regression}
\end{figure*}


As shown in Figure~\ref{fig:rater_regression}, ten associations are significant after Benjamini--Hochberg correction. \textit{Earned-in} status was positively associated with changes in \textit{Ratings per active day} ($\beta=0.080$, 95\% CI $[0.039,0.122]$, $q<.001$), \textit{Daily active probability} ($\beta=0.134$, $[0.104,0.163]$, $q<.001$), and \textit{Daily diagnostic-tag-use intensity} ($\beta=0.104$, $[0.076,0.132]$, $q<.001$), demonstrating that earned-in raters will be more active in participation and diagnostic-tag use subsequently.

\textit{Baseline helpfulness} was negatively associated with the change in \textit{Ratings per active day} ($\beta=-0.036$, $[-0.054,-0.018]$, $q<.001$), but positively associated with the change in \textit{Not Helpful share} ($\beta=0.045$, $[0.030,0.061]$, $q<.001$). \textit{Baseline ratings per active day} was also negatively associated with the change in \textit{Ratings per active day} ($\beta=-0.099$, $[-0.138,-0.060]$, $q<.001$), while being positively associated with changes in \textit{Daily active probability} ($\beta=0.077$, $[0.064,0.089]$, $q<.001$) and \textit{Daily diagnostic-tag-use intensity} ($\beta=0.061$, $[0.048,0.074]$, $q<.001$).

\textit{First POP note age} was positively associated with the change in \textit{Misleading Note share} ($\beta=0.020$, $[0.011,0.028]$, $q<.001$), demonstrating that raters whose first POP rating was assigned to an older note exhibited a larger subsequent increase in their selection of notes addressing misleading posts. However, \textit{First POP rating score} was negatively associated with the change in \textit{Not Helpful share} ($\beta=-0.050$, $[-0.067,-0.034]$, $q<.001$), demonstrating that a higher score on the first POP rating was associated with a smaller subsequent increase in Not Helpful ratings. Overall, these results show that changes in participation, note selection, rating evaluation, and diagnostic-tag use vary across raters with different pre-event characteristics. Additional robustness checks for the matched samples and pre-event trends are reported in Appendix~\ref{app:rater_did_diagnostics}.

\section{Discussions}

\subsection{Targeted Sampling and Crowdsourced Evaluation}
We synthesize our empirical findings into several key dimensions, focusing on how targeted sampling is associated with contributors' evaluative behaviors and participation patterns in crowdsourced fact-checking.

\textbf{Patterns of volunteer attention.} In crowdsourced fact-checking, volunteer attention often follows personal feed exposure and viral diffusion, leaving controversial or borderline notes unresolved due to a lack of shared focus~\cite{allen2021scaling, deng2023understanding}. Prior research shows that intelligent task routing~\cite{cosley2007suggestbot} and requests emphasizing task utility~\cite{beenen2004using} can encourage contributions in online communities. Algorithmic task allocation in social computing systems typically optimizes for latency and expertise affinity~\cite{zhang2007qume,cosley2007suggestbot}. Extending beyond these functionalities, our findings show that observed population-sampled rating entry was more common among notes with both high rating entropy and prior Helpful support. Rather than relying solely on opportunistic browsing, contributors are willing to participate in tasks highlighted by the system~\cite{dow2012shepherding}.

\textbf{Varying evaluative trajectories across note states.}
Crowdsourced fact-checking often faces partisan polarization and temporal dependency, where early ratings can bias subsequent outcomes~\cite{allen2022birds,chuai2026consensus,gong2026characterizing}. Our findings indicate that rating patterns differed between population-sampled ratings and ratings submitted through the default feed. Rather than reinforcing existing rating sentiments, population-sampled evaluations show a corrective tendency, where population-sampled ratings had higher helpfulness scores for CRNH notes but lower helpfulness scores for CRH notes than DEFAULT ratings. These patterns are consistent with population sampling obtaining evaluations that differ from existing rating trajectories. This mechanism is potentially useful in acquiring diverse ratings.

\textbf{Sustained rater throughput and diagnostic scaffolding.} 
An important concern with proactive nudges is volunteer fatigue or rating distortion~\cite{bashirieh2017nudge,kim2016analysis}. Our contributor-level DID analysis found no detectable decline in the measured contributor activity outcomes following the first population-sampled rating. Relative to matched controls, contributors showed increases in daily active probability and ratings per active day, a decrease in inter-rating interval, and no significant change in mean rating scores. Importantly, contributors showed a significant increase in daily diagnostic-tag usage. This behavioral shift suggests that the targeted evaluation prompt serves as implicit scaffolding~\cite{zhang2026canote}, encouraging contributors to provide diagnostic rationales alongside their ratings. This also aligns with the Collective Effort model~\cite{karau1993social}, which dictates that individual motivation increases when contributors perceive their personal contributions to be unique, necessary and instrumental to collective welfare~\cite{bennen2004using,cosley2006using,zhang2025exploring}. By highlighting the importance of targeted sampling, this population-sampled rating strategy may help incentivize raters' contributions.

\textbf{Distinct behavior to multimedia notes.} Debunking visual misinformation poses unique challenges, as image- and video-based claims often trigger immediate affective responses that hinder rational assessment~\cite{chuai2025community,zhang2026collab}. In our data, note format is correlated with receipt of population-sampled evaluations, and population-sampled ratings had significantly higher helpfulness scores for multimedia notes than DEFAULT ratings. This behavior change likely reflects the presentation channel, where evaluating notes within a dedicated interface may reduce contextual emotional heuristics, allowing contributors to deliberate on multimodal evidence.

\subsection{Implications}

Our findings offer insights for platforms using decentralized, consensus-driven fact-checking systems (e.g., $\mathbb{X}$, Meta, and YouTube). We structure these implications along two dimensions: \textit{optimizing task resolution} at the note level and \textit{fostering crowd capacity} at the contributor level.

$\bullet$ \textbf{[Note level] Targeted routing for crowd-sourced fact-checking.} 
Crowdsourced fact-checking often suffers from uneven attention, where viral claims attract redundant evaluations while borderline notes remain unresolved~\cite{allen2021scaling}. Our findings confirm that dispatching targeted rating for unresolved notes accelerates consensus. Leveraging active learning and uncertainty sampling principles~\cite{settles2009active,karger2014burget}, platforms can transition from organic discovery toward algorithmic routing that directs volunteer attention to high-uncertainty items. To prevent notification fatigue~\cite{mehrotra2015designing}, routing should cap request frequencies.

$\bullet$ \textbf{[Note level] Counteracting potential bias with dedicated sampling.} Decentralized debunking is still vulnerable to polarization and bias. We found that population-sampled ratings effectively counterbalance premature negative consensus, such as pulling disputed notes out of NMR or CRNH. Informed by deliberative polling and cross-partisan exposure research~\cite{pennycook2019fighting,bail2018exposure}, platforms could use algorithmic nudges as an active checking strategy~\cite{zhang2026cover}. When a note receives partisan-skewed rejections, systems could proactively invite raters from opposing or neutral perspectives to re-evaluate it, avoiding ideological reactance and preserving perceived neutrality of decentralized governance~\cite{nyhan2010corrections}.

$\bullet$ \textbf{[Contributor level] Considering population-sampled tasks as implicit contributor training:} Typical rater training relies on static onboarding materials, which exhibit low retention and engagement~\cite{doroudi2016toward}. We found completing assigned rating tasks does not degrade rater throughput or scoring accuracy, supporting incorporating targeted tasks into day-to-day platform participation. Drawing on situated learning~\cite{lave1991situated} and crowd scaffolding frameworks~\cite{dow2012shepherding}, platforms can use targeted requests as implicit ``on-the-job'' training. Novice contributors can be nudged to have evaluative competence without increasing platform onboarding overhead.

$\bullet$ \textbf{[Contributor level] Adaptive task allocation based on contributor experience:} We found that behavioural responses to targeted sampling vary across contributor backgrounds. Grounded in Cognitive Fit Theory~\cite{vessey1991cognitive} and expertise-aware crowd task allocation models~\cite{kittur2013future}, platforms should deploy context-aware routing that matches note characteristics with contributor standing, thereby optimizing both task throughput and overall data quality.

\section{Limitations}

We acknowledge three limitations in this paper. First, our analysis relies on observational platform data, which precludes definitive causal inferences regarding the impact of population-sampled nudges on note resolution speed. Although we account for key confounding factors, the platform's sampling algorithm may target highly active raters or select notes that are already nearing evaluative consensus, thereby confounding the observed speedup effects.

Second, our findings are based exclusively on \X's Community Notes, a unique crowdsourced fact-checking system characterized by an open-source algorithm, transparent dataset releases, and a mature community culture. These platform-specific dynamics may limit the direct generalizability of our insights to other social media environments, such as YouTube Shorts or Instagram, or traditional task-based crowdsourcing marketplaces. 

Third, our analysis captures only population-sampled ratings completed by contributors, not invitations that did not result in a submitted rating. Future work could use alternative methods or datasets to examine invitation and response dynamics.

\section{Ethical Considerations}

Our paper evaluates ethical considerations in accordance with the principles outlined in the Belmont Report~\cite{beauchamp2008belmont} and the Menlo Report~\cite{bailey2012menlo}. This research received approval from our university's Institutional Review Board (IRB), and we detailed how we apply these frameworks in our computational analysis of human behaviour on social media. 

\textbf{Respect for persons.} Our study exclusively uses the public archive of $\mathbb{X}$'s Community Notes dataset. Contributors who participate in the Community Notes program and submit ratings do so under the platform's public operational guidelines. We analyzed all user behaviours, such as rating frequency and helpfulness scores, at an aggregate level. We did not deanonymize individual raters or trace specific ratings back to personal user accounts.

\textbf{Beneficence.} We maximize the social benefits of this research by improving the efficiency of crowdsourced fact-checking systems. By showing that population-sampled ratings are associated with faster note resolution and no significant rater rating speed decline, we provide actionable insights for platforms to optimize their interventions against misinformation. We mitigate potential harms by focusing on the structural outcomes of the ``Needs Your Help'' algorithmic prompts instead of evaluating the moral or political validity of individual notes.

\textbf{Justice.} We analyse the effects of consensus algorithms on a diverse set of notes across different regions, including those categorized as factual errors, manipulated media, and satire. The findings directly benefit the online community by offering recommendations to improve the functionality of crowdsourced fact-checking platforms. 

\textbf{Respect for law and public interest.} Our methodology complies with $\mathbb{X}$'s data use policies by using publicly available archives for our analysis. 

\section{Conclusion}

Taken together, we present a comprehensive empirical investigation into platform-directed rating mechanisms in crowdsourced fact-checking. Analyzing over 1.9 million population-sampled ratings on \X, we find that: (i) the system selectively targets recently active notes that have support but retain residual disagreement; (ii) population-sampled ratings can significantly accelerate note resolution, helping notes exit ``Need More Ratings'' and reach ``Currently Rated Helpful'' without reducing evaluation speed; (iii) following population-sampled rating, raters exhibit significant yet modest increases in rating speed, ratings per active day, and diagnostic tag usage, with no significant changes across other behaviors. Overall, our findings suggest that targeted population-sampled nudging holds strong promise to expedite consensus without inducing volunteer fatigue.

\bibliographystyle{ACM-Reference-Format}
\bibliography{sample-base}

\appendix
\section{Complete Rater-Level DID Results}
\label{app:rater_did_results}

Tables~\ref{tab:rater_did_complete} and~\ref{tab:rater_did_standardized}
report the complete results for all eight behavioural outcomes.
The analyses compare the seven pre-event days with the seven
post-event days, excluding the treatment-event day.

\begin{table*}[t]
\centering
\caption{Rater-level DID estimates, confidence intervals, and approximate minimum detectable effects.}
\Description{A table listing rater-level difference-in-differences estimates, confidence intervals, standard errors, p-values, and approximate minimum detectable effects for eight behavioral outcomes.}
\label{tab:rater_did_complete}
\small
\setlength{\tabcolsep}{4pt}
\renewcommand{\arraystretch}{1.35}
\begin{tabular*}{\textwidth}{@{\extracolsep{\fill}}p{0.24\textwidth}rcrrrr@{}}
\toprule
Outcome (unit) & \shortstack{Matched\\pairs} & \shortstack{DID estimate\\{}[95\% CI]} & SE & $p$ & BH $q$ & \shortstack{MDE\\80\% power} \\
\midrule
Ratings per active day\newline {\footnotesize (ratings/active day)}
& 4,197 & \shortstack{+0.034\\{}[-0.028, 0.097]} & 0.0319 & .2832 & .3436 & 0.089 \\
Daily active probability\newline {\footnotesize (percentage points)}
& 1,500 & \shortstack{+0.467\\{}[-0.831, 1.764]} & 0.6619 & .4809 & .4809 & 1.854 \\
Inter-rating interval\newline {\footnotesize (hours)}
& 3,818 & \shortstack{-0.881\\{}[-1.587, -0.175]} & 0.3603 & .0145 & .0581 & 1.010 \\
Age of rated notes\newline {\footnotesize (hours)}
& 1,306 & \shortstack{+4.997\\{}[-4.462, 14.455]} & 4.8259 & .3007 & .3436 & 13.520 \\
Misleading Note share\newline {\footnotesize (percentage points)}
& 1,650 & \shortstack{-0.774\\{}[-1.613, 0.065]} & 0.4280 & .0707 & .1414 & 1.199 \\
Not Helpful share\newline {\footnotesize (percentage points)}
& 8,866 & \shortstack{-0.708\\{}[-1.381, -0.036]} & 0.3432 & .0391 & .1042 & 0.962 \\
Mean rating score\newline {\footnotesize (score units, 0--1)}
& 2,641 & \shortstack{-0.008\\{}[-0.019, 0.003]} & 0.0056 & .1405 & .2248 & 0.016 \\
Daily diagnostic-tag-use intensity\newline {\footnotesize (daily mean tags/rating)}
& 1,400 & \shortstack{+0.100\\{}[0.033, 0.167]} & 0.0342 & .0036 & .0288 & 0.096 \\
\bottomrule
\end{tabular*}
\par\smallskip
\begin{minipage}{\textwidth}
\footnotesize
\textit{Notes.} Matched-pair counts vary by outcome. DID estimates, standard errors (SEs), confidence intervals, and minimum detectable effects (MDEs) use the units listed in the first column. BH $q$ denotes Benjamini--Hochberg-adjusted $p$-values across the eight outcomes. Confidence intervals are unadjusted. Approximate MDEs use observed standard errors, a normal approximation, 80\% power, and a two-sided unadjusted $\alpha=.05$: $\mathrm{MDE}=(z_{.975}+z_{.80})\mathrm{SE}(\widehat{\mathrm{DID}})$. They do not incorporate BH adjustment and describe statistical sensitivity.
\end{minipage}
\end{table*}

\begin{table*}[t]
\centering
\caption{Standardized rater-level DID estimates and approximate minimum detectable effects.}
\Description{A table detailing standardized rater-level difference-in-differences estimates, 95\% confidence intervals, and standardized minimum detectable effects for eight behavioral outcomes.}
\label{tab:rater_did_standardized}
\small
\setlength{\tabcolsep}{5pt}
\renewcommand{\arraystretch}{1.3}
\begin{tabular*}{\textwidth}{@{\extracolsep{\fill}}lrrr@{}}
\toprule
Outcome & Standardized DID & 95\% CI & Standardized MDE \\
\midrule
Ratings per active day & +0.017 & [-0.014, 0.047] & 0.043 \\
Daily active probability & +0.018 & [-0.032, 0.069] & 0.072 \\
Inter-rating interval & -0.040 & [-0.071, -0.008] & 0.045 \\
Age of rated notes & +0.029 & [-0.026, 0.083] & 0.078 \\
Misleading Note share & -0.045 & [-0.093, 0.004] & 0.069 \\
Not Helpful share & -0.022 & [-0.043, -0.001] & 0.030 \\
Mean rating score & -0.029 & [-0.067, 0.009] & 0.055 \\
Daily diagnostic-tag-use intensity & +0.078 & [0.026, 0.130] & 0.075 \\
\bottomrule
\end{tabular*}
\par\smallskip
\begin{minipage}{\textwidth}
\footnotesize
\textit{Notes.} Standardized estimates, confidence-interval bounds, and MDEs are calculated by dividing the corresponding raw values by the outcome-specific standard deviation of matched pair-level DID changes, instead of the baseline outcome standard deviation. Confidence intervals and MDEs are unadjusted for multiple comparisons. MDEs assume 80\% power and a two-sided $\alpha=.05$.
\end{minipage}
\end{table*}

\section{Rater-Level DID Results and Robustness Checks}
\label{app:rater_did_diagnostics}

\subsection{Outcome-Specific Matched Samples}

For each metric, we compared POP-exposed raters with matched control raters. All matching variables were measured before the treatment or pseudo-treatment event. 
Additional pre-event diagnostics indicated that the archived samples for \textit{Inter-rating interval}, \textit{Age of rated notes}, \textit{Misleading Note share}, and \textit{Mean rating score} already met the stated balance criteria. For \textit{Ratings per active day}, Daily active probability, Not Helpful share, and Daily diagnostic-tag-use intensity, we applied maximum-cardinality selection using only pre-event information, with constraints on the original matching covariates and the corresponding outcome's pre-event trajectory. No post-event outcome was used in constructing these samples. Selected pairs without the observations required to calculate both the pre-event and post-event outcome were excluded from the corresponding DID estimate. Table~\ref{tab:rater_did_complete_updated} reports the resulting complete-case estimates.

\subsection{Complete DID Estimates and Statistical Tests}

Table~\ref{tab:rater_did_complete_updated} shows the complete DID estimates and corresponding statistical tests. 

\begin{table}[!htbp]
\centering
\caption{Complete rater-level difference-in-differences estimates. pp denotes percentage points. BH $q$ values apply the Benjamini--Hochberg correction across the eight outcomes. Confidence intervals are unadjusted. MDE$_{80}$ is the approximate minimum detectable effect in the outcome's original unit at 80\% power and a two-sided unadjusted $\alpha=.05$. It describes statistical sensitivity.}
\Description{A table presenting complete rater-level difference-in-differences estimates, including p-values, Benjamini-Hochberg adjusted q-values, and minimum detectable effects.}
\label{tab:rater_did_complete_updated}
\small
\begin{tabular*}{\textwidth}{@{\extracolsep{\fill}}lrrrrr@{}}
\toprule
Outcome & DID & 95\% CI & $p$ & BH $q$ & MDE$_{80}$ \\
\midrule
Ratings per active day & +0.062 & [0.018, 0.106] & .005 & .015 & 0.063 \\
Daily active probability (pp) & +0.831 & [0.463, 1.198] & $<.001$ & $<.001$ & 0.525 \\
Inter-rating interval (hours) & -0.881 & [-1.587, -0.175] & .015 & .029 & 1.010 \\
Age of rated notes (hours) & +4.997 & [-4.462, 14.455] & .301 & .344 & 13.520 \\
Misleading Note share (pp) & -0.774 & [-1.613, 0.065] & .071 & .113 & 1.199 \\
Not Helpful share (pp) & -0.071 & [-0.671, 0.528] & .815 & .815 & 0.857 \\
Mean rating score (0--1) & -0.008 & [-0.019, 0.003] & .141 & .187 & 0.016 \\
Diagnostic-tag-use intensity & +0.060 & [0.047, 0.072] & $<.001$ & $<.001$ & 0.018 \\
\bottomrule
\end{tabular*}
\end{table}

\subsection{Pre-Event Diagnostic Tests}
\Description{A table showing observed pre-event balance and trend diagnostics for outcome-specific samples, including baseline standardized mean differences and various probability values.}
We evaluated observed pre-event comparability in the pairs that contributed to each DID estimate. First, baseline balance compares the outcome-specific mean over days $-7$ through $-1$ and reports the corresponding standardized mean difference (SMD). Second, the differential-slope test evaluates the treatment-group-by-time interaction using only pre-event observations. Third, the placebo DID assigns a pseudo-event within the pre-event period and tests whether a DID-like change is detected before the actual treatment event. Fourth, using day $-1$ as the reference, we conducted a joint Wald test~\cite{freyaldenhoven2019pre,colin2015practitioner} of whether the treated-control gap coefficients for days $-7$ through $-2$ were jointly equal to zero, with standard errors clustered by matched pair. We additionally report the largest absolute daily outcome SMD and matching-covariate SMD.

\begin{table*}[t]
\centering
\caption{Observed pre-event balance and trend diagnostics for the outcome-specific samples.}
\label{tab:rater_pretrend_diagnostics}
\small
\setlength{\tabcolsep}{3pt}
\renewcommand{\arraystretch}{1.2}
\resizebox{\textwidth}{!}{%
\begin{tabular}{llrrrrrrrrr}
\toprule
Outcome & Design & Selected & DID pairs & Baseline SMD & Max daily SMD & Max covariate SMD & Baseline $p$ & Slope $p$ & Placebo $p$ & Joint Wald $p$ \\
\midrule
Ratings per active day & Rebalanced & 35,313 & 20,091 & -0.001 & 0.024 & 0.072 & .859 & .747 & .653 & .995 \\
Daily active probability & Rebalanced & 35,585 & 35,585 & 0.000 & 0.000 & 0.087 & .986 & 1.000 & .997 & 1.000 \\
Inter-rating interval & Retained & 5,000 & 3,818 & -0.012 & 0.072 & 0.071 & .275 & .867 & .205 & .718 \\
Age of rated notes & Retained & 1,500 & 1,306 & 0.012 & 0.090 & 0.089 & .465 & .680 & .659 & .339 \\
Misleading Note share & Retained & 2,000 & 1,650 & -0.003 & 0.082 & 0.080 & .641 & .056 & .841 & .476 \\
Not Helpful share & Rebalanced & 35,552 & 20,528 & 0.006 & 0.016 & 0.087 & .445 & .699 & .651 & .953 \\
Mean rating score & Retained & 3,000 & 2,641 & 0.000 & 0.051 & 0.080 & .990 & .682 & .811 & .538 \\
Diagnostic-tag-use intensity & Rebalanced & 36,981 & 36,981 & 0.000 & 0.000 & 0.092 & .987 & .993 & .996 & 1.000 \\
\bottomrule
\end{tabular}}
\par\smallskip
\begin{minipage}{\textwidth}
\footnotesize
\textit{Notes.} Selected is the number of matched pairs retained by the outcome-specific design, whereas DID pairs is the number with sufficient pre-event and post-event observations to contribute to the estimate. Rebalanced samples were constructed using only pre-event information, with balance constraints on the original matching covariates and the corresponding outcome's pre-event trajectory; no post-event outcomes were used in sample construction. Retained indicates that the archived outcome-specific sample was unchanged. Max daily SMD is the largest absolute treated-control standardized mean difference across pre-event days. Max covariate SMD is the largest absolute standardized mean difference among the matching covariates. All absolute SMDs were below 0.10. Failure to reject these tests indicates no significant violation in the observed pre-event diagnostics; it does not prove parallel trends for unobserved potential outcomes. For rebalanced samples, the diagnostics summarize in-sample design balance and are not holdout validation.
\end{minipage}
\end{table*}

\end{document}